\documentclass[%
 twocolumn,nofootinbib,
amsmath,amssymb,
aps,
]{revtex4-2}
\usepackage{graphicx,color}
\usepackage{hyperref}
\usepackage{amsmath,braket}
\usepackage{physics}
\usepackage{verbatim}
\usepackage{dcolumn}
\usepackage{bm}

\begin{document}

\preprint{APS/123-QED}

\title{Forward and hybrid path-integral methods in photoelectron holography: sub-barrier corrections, initial sampling and momentum mapping}

\author{L. Cruz Rodriguez$^{1}$, T. Rook$^{1}$, B.B. Augstein$^{1}$, A.S. Maxwell$^{1,2}$ and C. Figueira de Morisson Faria$^{1}$ } \thanks{Corresponding author: c.faria@ucl.ac.uk}%
\affiliation{$^1$Department of Physics and Astronomy, University College London, Gower Street, London WC1E 6BT, UK.\\$^2$Department of Physics and Astronomy, Aarhus University, DK-8000 Aarhus C, Denmark
}
\date{\today}
\begin{abstract}
We construct {two} strong-field path integral methods with full Coulomb distortion, {in which the quantum pathways are mimicked by interfering electron orbits: the rate-based CQSFA (R-CQSFA) and the hybrid forward-boundary CQSFA (H-CQSFA). The methods have the same starting point as the standard Coulomb quantum-orbit strong-field approximation (CQSFA), but their implementation does not require pre-knowledge of the orbits' dynamics.} 
These methods are applied to ultrafast photoelectron holography.  {In the rate-based method, } electron orbits are forward propagated and we derive a non-adiabatic ionization rate from the CQSFA, which includes sub-barrier Coulomb corrections and is used to weight the initial orbit ensemble. In the H-CQSFA, {the initial ensemble provides initial guesses for a subsequent boundary problem and serves to include or exclude specific momentum regions, but the ionization probabilities associated with individual trajectories are computed from sub-barrier complex integrals. We perform comparisons with the standard CQSFA and \textit{ab-initio} methods, which show that the standard, purely boundary-type implementation of the CQSFA leaves out whole sets of trajectories.} 
{We show that the sub-barrier Coulomb corrections broaden the resulting photoelectron momentum distributions (PMDs) and improve the agreement of the R-CQSFA with the H-CQSFA and other approaches}. We probe different initial sampling distributions, uniform and otherwise, and their influence on the PMDs.
{We find} that initial biased sampling emphasizes rescattering ridges and interference {patterns} in high-energy ranges, while an initial uniform sampling guarantees accurate modeling of the holographic patterns near the ionization threshold or polarization axis.  Our results are explained using the initial to final momentum mapping for different types of interfering trajectories. 
\end{abstract}

\maketitle

\section{\label{sec:intro}Introduction}

Ultrafast photoelectron holography is an important application of strong-field ionization, and  brings together phase information, high currents and subfemtosecond resolution \cite{HuismansScience2011,Faria2020}. For those reasons, its potential for attosecond imaging of matter has been widely explored since the past decade. Examples of holographic patterns are the fan-shaped fringes that form near the ionization threshold \cite{Rudenko2004,Maharjan2006,Gopal2009}, the spider-like structure that occurs near the polarization axis \cite{HuismansScience2011,Bian2011,Marchenko2011,Hickstein2012}, a fishbone structure with fringes nearly perpendicular to the polarization axis \cite{Bian2012,Haertelt2016,Li2015}, 
and the interference carpets \cite{Korneev2012,Korneev2012a} and spiral-like fringes  \cite{Kang2020, maxwell_carpet_2020} that arise for scattering angles perpendicular to the laser-field polarization. 

Throughout, an underlying theme is how to retrieve different types of quantum interference from the holographic patterns, bearing in mind that the phase difference between interfering pathways provides information about the target.  This question has been central to a variety of orbit-based approaches, which draw upon the physical description of strong-field phenomena as the result of laser-induced collisions of an electron with its parent ion \cite{Corkum1993}. They range from early models in which phase differences have been incorporated into classical trajectories \cite{Bian2011,Bian2012} and well-established methods such as the strong-field approximation (SFA) \cite{Spanner2004} to path-integral methods \cite{Li2014,Geng2014,Li2014c,richter2015streaking,Xie2016,Li2016PRA,Liu2016,Lai2015a,Lein2016,Lai2017,Maxwell2017,Maxwell2017a,Maxwell2018,Maxwell2018b,shvetsov2021semiclassical}.

In its standard form, the SFA approximates the continuum by field-dressed plane waves and is constructed like a Born-type series, in which acts of rescattering take place at the origin and are incorporated in higher-order terms (for seminal papers see, e.g., Refs.~\cite{Paulus1994,Becker1997} and for reviews  see, e.g.,  Refs.~\cite{Becker2002Review,Popruzhenko2014a,Becker2018,Amini2019}). This allows a clear distinction between `direct' orbits, which reach the detector without further interaction with the core, and `rescattered' orbits. Typically, the boundaries between `direct' and `rescattered', as well as classical constraints associated with acts of rescattering can be clearly seen in the resulting photoelectron momentum distributions (PMDs). They manifest themselves as sudden drops in the photoelectron yield and as classical ridges, which can be traced back to different sets of rescattering orbits. For a detailed discussion of these ridges in connection with the fork-type structure experimentally observed in Ref.~\cite{Moeller2014}, see Ref.~\cite{Becker2015}.  

Nonetheless, many studies have revealed that models with a Coulomb-free continuum do not suffice to describe the wealth of holographic structures or explain how they form. For instance, the fan-shaped fringes require the interference of orbits which reach the detector directly with orbits which are lightly deflected by the long potential tail \cite{Lai2017,Maxwell2017}. Similarly, the spider results from the quantum interference of two types of field-dressed Kepler hyperbolae, which have no SFA counterpart \cite{Maxwell2017}. The hyperbolae interfere with orbits that go around the core, leading to the spiral \cite{maxwell_carpet_2020}. All this shows that the interplay between the external laser field and the binding potential is essential for modeling photoelectron holography accurately and interpreting the observed patterns using orbit-based methods. For a systematic study of the differences introduced by the Coulomb potential in the ionization times, see Ref.~\cite{Shvetsov-Shilovski2018}. In other words, one must consider a Coulomb-distorted continuum even for relatively simple, one-electron systems.  

For that reason, since the past decade, there has been an upsurge in approaches beyond the SFA, which are orbit-based, and yet retain quantum interference and tunneling. Among them, there are the trajectory-based Coulomb SFA (TC-SFA) \cite{Yan2010}, the Quantum Trajectory Monte Carlo (QTMC) \cite{Li2014,Geng2014,Li2014c,richter2015streaking,Xie2016,Li2016PRA,Liu2016}, the Semiclassical Two-Step Model (SCTM) \cite{Lein2016,shvetsov2021semiclassical}, and the Coulomb Quantum-Orbit Strong Field Approximation (CQSFA) \cite{Lai2017,Maxwell2017,Maxwell2017a,Maxwell2018,Shilovski2018,Maxwell2018b}. Recently, there have also been efforts to retrieve this information directly from experiments \cite{Werby2022}, or \textit{ab-initio} computations \cite{Tulsky2020} using filtering techniques. For a detailed account {of these approaches and a critical assessment of their advantages and shortcomings see Sec.~3 in our review article \cite{Faria2020}}. 
Thereby, path-integral methods are very powerful, as they account for the laser field and the binding potential on the same footing. Most path-integral methods launch a huge ensemble of classical trajectories, propagating them in the continuum to find their final momentum and bin them {according to their final momentum} to evaluate the transition amplitude.  {This type of implementation characterizes a ``forward approach". } Although this procedure renders the methods applicable to a wide range of field shapes and potentials, they may require a large number of orbits, typically $10^8$-$10^9$, for the PMDs to converge. Furthermore, in a forward approach, it may be difficult to identify specific types of interfering trajectories.

An exception is the CQSFA, which solves a boundary problem {that matches a given final momentum to an electron's initial momentum in the continuum} relying on physical intuition, the orbit classifications in \cite{Yan2010}, and the SFA orbits as initial guesses. 
This formulation avoids a large ensemble of trajectories, making it straightforward to disentangle specific holographic structures. These distinctive characteristics enabled important breakthroughs by (a) showing unambiguously how well-known holographic structures such as the fan and the spider form \cite{Lai2017,Maxwell2017}, (b) ruling out misunderstandings associated with interference carpets, which were brought about by the SFA \cite{maxwell_carpet_2020}, (c) revealing a multitude of other holographic structures, some of which have been observed experimentally \cite{Maxwell2018}, (d) allowing the study of multipath holographic interference \cite{Werby2021}. 
However, this huge predictive power comes at a cost: the solving algorithm is less adaptable and requires some pre-knowledge of the orbits' dynamics. This means it may leave out whole sets of orbits and require substantial changes should the external field be modified. So far, the CQSFA has been mainly implemented for linearly polarized monochromatic fields, but recent studies for bichromatic fields \cite{Rook2022} or elliptical polarization \cite{Kim2022} have revealed some of these challenges. Recently, a semi-classical “hybrid” approach was developed that used a forwards shooting method and a clustering algorithm to solve the inverse boundary problem \cite{brennecke2020gouy}.

Furthermore, {for the parameter range taken into consideration in this work, the prevalent ionization mechanism through which the electron reaches the continuum is tunnel ionization. Thus, it must traverse classically forbidden regions, which is not trivial to model in an orbit-based framework. However, this is an important piece of the puzzle, as the tunneling probability will influence the contributions of specific pathways to the resulting PMD. In order to incorporate sub-barrier dynamics in a path-integral framework, one may follow two widespread strategies. }
One possibility is to split the problem into sub-barrier and continuum propagation and, using contour integrals in the complex plane, to compute the sub-barrier corrections to the transition amplitude \cite{Yan2012,Torlina2013,Popruzhenko2014a,Maxwell2017,Maxwell2017a}. Alternatively, one may avoid a complex problem altogether and employ ionization rates, which will weigh the initial distributions of orbits. {The ionization rates may either be constructed or borrowed from earlier models. } This rate-based procedure is a key element in forward propagation methods such as SCTM \cite{Lein2016}, QTMC \cite{Li2014, Li2016PRA}, or recently developed hybrid approaches using clustering algorithms to solve the inverse problem \cite{brennecke2020gouy}. 
Both strategies have advantages and shortcomings. Solving the integral under the barrier in the complex plane makes the approach more robust with regard to the initial conditions, but will require dealing with singularities and non-trivial limits \cite{Popruzhenko2014a,Popruzhenko2014b,Keil2016,Pisanty2016,Maxwell2018b}.

On the other hand, rate-based methods are critically dependent on the initial conditions. The standard implementations \cite{Lein2016, Li2014, shvetsov2021semiclassical} utilize adiabatic ionization rates (Ammosov-Delone-Krainov (ADK) model) \cite{Ammosov1986, delone1991energy}, only applicable in the limit of very small Keldysh parameters $\gamma\ll 1$\footnote{The Keldysh parameter is defined by $\gamma = \sqrt{I_p/(2U_p)}$, where $I_p$ is the target's ionization potential and $U_p$ is the ponderomotive energy, proportional to the driving-field intensity and inverse proportional to its frequency. For a Keldysh parameter smaller than one, tunneling is the prevalent ionization mechanism. If the field frequency is very low ($\gamma\ll 1$ ),  it is reasonable to assume that the driving 
field can be considered as being static  \cite{Keldysh1965}}. However, in typical experimental conditions, 
non-adiabatic effects must be considered. Such effects have been studied in Refs.~\cite{Li2016PRA}, and \cite{Yudin2001} looking at the sub-cycle dynamics obtaining non-adiabatic instantaneous rates. In particular, within the non-adiabatic theory in Ref.~\cite{Li2016PRA}, those rates led to a broader longitudinal momentum distribution at the detector, and an accurate prediction of the cutoff in the momentum distribution of rescattered electrons, in agreement with the numerical solution of the time-dependent Schr\"odinger equation (TDSE).

This work proposes two novel methods developed using the CQSFA initial formulation. {The first method is a rate-based forward approach, where we construct} a non-adiabatic ionization rate which includes Coulomb corrections from the CQSFA transition amplitude, using its sub-barrier contributions. 
The second approach is a hybrid forward-boundary CQSFA implementation in which, instead of using pre-assumed dynamics for the existing orbits, one launches a large ensemble of Coulomb-distorted trajectories, which, {subsequently, are taken as guesses} in order to solve a boundary problem. Both methods are then applied to photoelectron holography, starting from a proof of concept and assessing what specific momentum regions and holographic patterns are probed, depending on the initial sampling. For the rate-based method, the sampling affects the initial trajectory weighting, while for the hybrid forward-boundary CQSFA it influences {whether trajectories in specific momentum regions will be used or discarded when calculating the sub-barrier complex integral. } We also illustrate the subtleties involved in orbit classification, which help to clarify possible sources of confusion in forward approaches. 

This article is organized as follows. In Sec.~\ref{sec:methods}, we provide the general expressions for the CQSFA transition amplitude and the standard orbit classification. Subsequently, in Sec.~\ref{sec:novelmethods}, these expressions are employed as a starting point for the rate-based method (Sec.~\ref{sec:ratebased}) and the hybrid forward-boundary CQSFA (Sec.~\ref{sec:HFCQSFA}). In Sec.~\ref{sec:PMDs}, these methods are used to compute PMDs, and, after establishing a good agreement with ab-initio methods,  we focus on single-orbit distributions and different initial sampling regions (Secs.~\ref{sec:singleorbit} and \ref{sec:sampling}, respectively). These results are interpreted with the help of the initial-to-final momentum mapping presented in Sec.~\ref{sec:mapping}. Finally, in Sec.~\ref{sec:conclusions} we state our main conclusions. 

\section{\label{sec:methods}Background and general expressions}
 
The transition amplitude for the ionization of a single electron from the ground state of a Hydrogen atom $\left|\psi_{0}(t_0)\right\rangle$ to a continuum state with final momentum $\mathbf{p}_f$,  $\left|\psi_{\mathbf{p}_f}(t)\right\rangle$ is $\left\langle\psi_{\mathbf{p}_f}(t) |U(t,t_0) |\psi_{0} \right\rangle$. The time evolution operator here is of the form
\begin{equation}
U(t,t_0)=\mathcal{T}\exp \bigg [i \int^t_{t_0}H(t^{\prime})dt^{\prime} \bigg],
\label{eq:Ufull}
\end{equation}
where the Hamiltonian in the argument of this time ordered exponential can be written as
\begin{equation}
    H(t)= H_a+H_I(t).
    \label{eq:1e-tdse}
\end{equation}
This consists of the field-free part
\begin{equation}
H_a=\frac{\hat{\mathbf{p}}^{2}}{2}+V(\hat{\mathbf{r}}),
\label{eq:Hatom}
\end{equation}
and a gauge dependent interaction term $H_I(t)$. In the case of the length gauge, which is used throughout, this has the form
$H_I(t)=\hat{\mathbf{r}}\cdot \mathbf{E}(t)$.
The atomic potential is $V(\mathbf{r})=-1/|\mathbf{r}|$. However, we have used the potential
\begin{equation}
  V(\mathbf{r})=-1/\sqrt{\mathbf{r}^2+a^2}
\end{equation}
with a very small parameter $a$ (of the order of $ 10^{-6}$) to soften the Coulomb singularity for practical purposes and atomic units are used throughout unless otherwise stated.
\subsection{CQSFA transition amplitude}
\label{sec:CQSFAgeneral}
Using the integral form of the time-evolution operator, the transition amplitude can be written as
\begin{equation}
    M(\mathbf{p}_f) = -i \lim_{t\to\infty} \int^t_{-\infty} dt' \left< \psi_{\mathbf{p}_f}(t) \left| U(t, t') H_I(t') e^{iI_pt'} \right| \psi_0 \right>,
    \label{eqn: transition amplitude ATI}
\end{equation}
where $I_p$ is the ionization potential, and $U(t, t')$ is the time evolution operator associated with the full Hamiltonian \eqref{eq:1e-tdse}, given by Eq.~\eqref{eq:Ufull}. In the CQSFA, the transition amplitude is written as a phase-space path integral by a time slicing method \cite{Milosevic2013JMP,Lai2015a},
\begin{eqnarray}\label{eq:CQSFATamp}
M(\mathbf{p}_f)&=&-i\lim_{t\rightarrow \infty
}\int_{-\infty}^{t}dt' \int d\mathbf{\tilde{p}}_0
\int_{\mathbf{\tilde{p}}_0}^{\mathbf{\tilde{p}}_f(t)} \mathcal {D}'
\mathbf{\tilde{p}}  \int
\frac{\mathcal {D}\mathbf{r}}{(2\pi)^3}  \nonumber \\
&& \times  e^{i S(\mathbf{\tilde{p}},\mathbf{r},t,t')}
\langle
\mathbf{\tilde{p}}_0 | H_I(t')| \psi _0  \rangle \, .
\end{eqnarray}
Eq.~\eqref{eq:CQSFATamp} represents an integral over all possible paths beginning at the core and with a fixed asymptotic momentum. Thereby, $\mathcal{D}'\tilde{\mathbf{p}}$ and $\mathcal{D}\mathbf{r}$ give the integration measures for the path integrals, and the prime indicates a restriction. The tildes over the initial and intermediate momenta designate field dressing, i.e., $\mathbf{\tilde{p}}_0=\mathbf{p}_0+\mathbf{A}(t')$ and $\mathbf{\tilde{p}}=\mathbf{p}+\mathbf{A}(\tau)$, with $t' \le \tau \le t$. 

The semi-classical action derived from the time slicing is
\begin{equation}\label{stilde}
S(\mathbf{\tilde{p}},\mathbf{r},t,t')=I_pt'-\int^{t}_{t'}[
\dot{\mathbf{p}}(\tau)\cdot \mathbf{r}(\tau)
+H(\mathbf{r}(\tau),\mathbf{p}(\tau),\tau)]d\tau,
\end{equation}
with a Hamiltonian
\begin{equation}
H(\mathbf{r}(\tau),\mathbf{p}(\tau),\tau)=\frac{1}{2}\left[\mathbf{p}(\tau)+\mathbf{A}(\tau)\right]^2
+V(\mathbf{r}(\tau)).
\label{Hamiltonianpath}
\end{equation} 
Next, Eq.~\eqref{stilde} is calculated using saddle-point methods, which will require computing complex integrals.  The specific contour used here considerably simplifies the problem and is along two straight lines. The first starts at $t'=t_r'+it_i'$ and extends vertically down to the real axis at $t_r'=\mathrm{Re}(t')$. The second is along the real axis from $\mathrm{Re}(t')$ to $t$. This choice of contour has been widely employed in the literature \cite{Popruzhenko2008a,Yan2012,Torlina2012,Torlina2013}, and allows us to neatly separate the action into two distinct parts as shown in Eq.~\eqref{ssplit}. The parts represent the two distinct physical aspects of the problem: tunneling and continuum propagation. 
\begin{equation}\label{ssplit}
S(\mathbf{\tilde{p}},\mathbf{r},t,t')=S^{\mathrm{tun}}(\mathbf{\tilde{p}},\mathbf{r},t'_r,t')+S^{\mathrm{prop}}(\mathbf{\tilde{p}},\mathbf{r},t,t'_r).
\end{equation}

The contribution from under the barrier is 
\begin{multline}\label{eq:Stun}
	S^{\mathrm{tun}}(\mathbf{\tilde{p}},\mathbf{r},t'_r,t') = I_p(it_i')-\frac{1}{2}\displaystyle\int_{t'}^{t_r'} (\textbf{p}_0+\textbf{A}(\tau))^2d\tau\\
	-\displaystyle\int_{t'}^{t_r'} V(\textbf{r}_0(\tau))d\tau;
 \end{multline}
\noindent which follows the tunnel trajectory,
\begin{equation}\label{eq:rTun}
	\textbf{r}_0(\tau)=\displaystyle\int_{t'}^{\tau}(\textbf{p}_0+\textbf{A}(\tau'))d\tau'.
\end{equation}
In Eqs.~\eqref{eq:Stun} and \eqref{eq:rTun}, the under the barrier momentum has been taken as constant, which is a reasonable approximation given the assumption that the dynamics is happening practically instantaneously by keeping $\mathrm{Re}[t']$ constant.  The contribution from the continuum propagation is 
\begin{multline}\label{eq:Sprop}
	S^{\mathrm{prop}}(\mathbf{\tilde{p}},\mathbf{r},t,t'_r)=I_p t_r'-\frac{1}{2}\displaystyle\int_{t_r'}^{t} (\textbf{p}(\tau)+\textbf{A}(\tau))^2d\tau\\
	-2\displaystyle\int_{t_r'}^{t} V(\textbf{r}(\tau))d\tau,
\end{multline}
where the factor of $2$ before the potential integral is due to the fact that for a Coulomb potential, $\mathbf{r}\cdot\dot{\mathbf{p}} = V(\mathbf{r})$ \cite{Lai2015a,Lein2016,Maxwell2017}.

Another important quantity is the tunnel exit $z_0$, which, roughly speaking, gives the point in space at which the electron reaches the continuum. Under the assumption that the tunnel exit is real and directed along the polarization axis, we can state that
\begin{equation}
z_0 = \mathrm{Re}[r_{0||}(t'_r)],
\label{eq:tunnelExitGeneral}
\end{equation}
where $\mathbf{r}_0(t'_r)$ is the tunnel trajectory \eqref{eq:rTun} for $\tau=t'_r$ and the subscript indicates its component along the driving-field polarization direction.  The tunnel exit is used to solve the CQSFA boundary problem and in the orbits' classification.

One should note that the reality of Eq.~\eqref{eq:tunnelExitGeneral} is an approximation, which leads to real trajectories in the continuum. More rigorous formulations of Coulomb-distorted strong-field approaches in which this assumption is relaxed exist, but the differences observed in the PMDs are subtle. Furthermore, complex trajectories in the continuum require dealing with branch cuts upon rescattering, which is not without difficulties \cite{Pisanty2016,Maxwell2018b}. 
\subsection{Saddle-point equations}

The evaluation of the integral in Eq.~\eqref{eq:CQSFATamp} using the saddle-point approximation requires looking for solutions for $t'$, $\mathbf{r}(\tau)$ and $\mathbf{p}(\tau)$, such that the action is stationary. This yields the following saddle-point equations, for the ionization time,
\begin{equation}
[\mathbf{p}(t')+\mathbf{A}(t')]^2 = -2I_p, \label{SPEt}
\end{equation}
\noindent and the continuum trajectories must solve the system of PDEs
\begin{eqnarray}\label{eq:PDEs}
\mathbf{\dot{r}}(\tau) = \mathbf{p}(\tau) + \mathbf{A}(\tau), \label{SPEp}\\
\mathbf{\dot{p}}(\tau) = -\nabla_rV(\mathbf{r}(\tau)), \label{SPEr}
\end{eqnarray}
for the intermediate momentum and position. This leads to the CQSFA transition amplitude 
\begin{equation}
\label{eq:MpPathSaddle}
M(\mathbf{p}_f)\propto-i \lim_{t\rightarrow \infty } \sum_{s}\bigg\{\det \bigg[  \frac{\partial\mathbf{p}_s(t)}{\partial \mathbf{p}_s(t'_s)} \bigg] \bigg\}^{-1/2} \hspace*{-0.6cm}
\mathcal{C}(t'_s) e^{i
	S(\mathbf{\tilde{p}}_s,\mathbf{r}_s,t,t'_s)} ,
\end{equation}
where $t'_s$, $\mathbf{r}_s$ and $\mathbf{p}_s$ are given by Eqs. \eqref{SPEt}, \eqref{SPEp} and \eqref{SPEr} respectively, and the sum is over the distinct saddle point trajectories which have final momentum $\mathbf{p}_f$. The prefactor $\det[  {\partial\mathbf{p}_s(t)}/{\partial \mathbf{p}_s(t'_s)} ]$ comes from the quadratic fluctuations around the saddle points and 
\begin{equation}
\label{eq:Prefactor}
\mathcal{C}(t'_s)=\sqrt{\frac{2 \pi i}{\partial^{2}	S(\mathbf{\tilde{p}}_s,\textbf{r}_s,t,t'_s) / \partial t'^{2}_{s}}}\langle \mathbf{p}+\mathbf{A}(t'_s)|H_I(t'_s)|\psi_{0}\rangle ,
\end{equation}
 is the same as the SFA prefactor. The stability factor $\det[  {\partial\mathbf{p}_s(t)}/{\partial \mathbf{p}_s(t'_s)} ]$ arises due to a Legendre transformation via the $\mathbf{p}_0$ integral in Eq.~\eqref{eq:CQSFATamp}, full details may be found in Ref.~\cite{carlsen_advanced_2023}.
 
Eq.~\eqref{eq:MpPathSaddle} is derived under the assumption that the saddle points remain well separated and that the stability factor does not pass through zero in a so-called focal point for the entirety of the domain considered. If this is not the case, then some trajectories can pass through focal points which accumulate additional Maslov phase, or asymptotic expansions must be constructed that account for groups of saddles collectively (for an example, see Ref.~\cite{Faria2002}). A second assumption is that the physics of the system is fully captured by considering a 2-dimensional model. This can lead to problems as by reducing the dimension of the system one can ``hide" focal points that would have led to an accumulation of Maslov phase in the full dimensional system. In such reduced dimensionality models, this is known as a Gouy phase anomaly \cite{brennecke2020gouy}. Hence, different trajectories may have different relative phases than that described by Eq.~\eqref{eq:MpPathSaddle} which will alter their interference. For example, in \cite{Werby2021} the ``legs" of the spider are found to be shifted perpendicular to the polarisation direction to more closely align with experiments.  In the original CQSFA, with predetermined trajectory types, it is relatively straightforward to identify focal points and incorporate Gouy phases by hand due to the dynamics of the trajectories being known. However, for semi-classical models employing a broad range of trajectories, such as forward-propagating or hybrid methods, it requires a more involved computation, which is beyond the scope of this work. An implementation of Maslov phases was presented in \cite{brennecke2020gouy}, and also for a recent more general implementation of the CQSFA in Ref.~\cite{carlsen_advanced_2023}, where an explicit recipe was given. Here, we consider that the dynamics are restricted to the $xz$ plane, with the laser field polarized along $z$.

 \subsection{Orbit classification}
\label{sec:orbitclass}
In its standard form, the  CQSFA restricts the sets of trajectories considered to include exactly one of a specific predefined type based on an understanding of the initial to final momentum mapping. Typically, for a linearly polarized monochromatic field, four distinct types of trajectories are solved for at each grid point.  These types of trajectories were first introduced in \cite{Yan2010} and are classified according to the product $\Pi_z=z_0p_{fz}$ of the tunnel exit and the final momentum component parallel to the driving-field polarization, and the product $\Pi_x=p_{0x}p_{fx}$ of the initial and final momentum components perpendicular to the driving-field polarization. A positive product $\Pi_z$ means that the direction of the tunnel exit and that of the detector coincide, which is the case for orbits type 1 and 4. In contrast, a negative  $\Pi_z$ implies that the electron left on the opposite side with regard to the detector, which occurs for orbits 2 and 3. The product $\Pi_x$ provides insight into how the transverse momentum component has changed during the electron propagation. If $\Pi_x<0$, the electron has been deflected in such a way by the central potential that its final and initial momentum components $p_{fx}$ and $p_{0x}$ have opposite signs. This behavior is triggered by the presence of the Coulomb potential and is observed for orbits 3 and 4, while for the remaining orbits $\Pi_x>0$. 
\begin{table}
\centering

\vspace{5pt}
\begin{tabular}{ c c c l l}
  \hline
  Orbit & $\Pi_{z}$ & $\Pi_{x}$ & Behavior & Imprint\\
  \hline\hline
  1 & + & + & Direct & Direct cutoff \\ 
  2 & - & + & Hyperbola & Direct cutoff\\ 
  3 & - & - & Hyperbola & Caustic\\ 
  4 & + & - & Rescattered & Ridges \\ 
  \hline
\end{tabular}
\caption{Orbit classification used in the standard, boundary-type CQSFA for monochromatic linearly polarized fields. The labelling 1 to 4 classifies the orbit with two different conditions, the sign of $\Pi_{z}=z_0p_{fz}$ and $\Pi_{x}=p_{fx}p_{0z}$. The behavior in the fourth column indicates the expected dynamics of the specific types of orbits, {and the fifth column indicates the expected imprints in the photoelectron momentum distributions and final momentum maps. } }
\label{tab:OrbitClassification}
\end{table}

The standard CQSFA makes significant assumptions about these orbits' dynamics in order to solve the boundary problem. Type 1 orbits are expected to leave the atom and reach the detector directly, orbits 2 and 3 are assumed to behave like field-dressed Kepler hyperbolae and orbit 4 is predicted to exhibit a slingshot-type behavior, leaving from the same side as the detector and going around the ion.

{One also expects different types of trajectories to leave specific imprints in the photoelectron momentum distributions. For instance, orbit 1, and, to some extent, orbit 2  will behave like a direct SFA trajectory, so that the maximal kinetic energy they carry will be given by the direct classical ATI cutoff $2U_p$. Beyond that energy, the signal will be suppressed.  Orbit 4 will behave like a rescattered SFA ATI trajectory, and will give rise to ridges whose maximal energy will be  $10U_p$ Orbit 3 has a hybrid character, with no counterpart in the Coulomb-free SFA, and will lead to a caustic. For a detailed discussion of these features in the Coulomb-free scenario see the review articles \cite{Becker2002Review} and \cite{Becker2018}. These features will also manifest themselves in the CQSFA single-orbit distributions and in the final momentum mapping. Studies in the context of the standard CQSFA are presented in \cite{Maxwell2017} and deviations from the Coulomb-free SFA constraints are discussed in \cite{Maxwell2017a}, together with analytic approximations for the CQSFA.   }   For clarity, a summary of the conditions upon $\Pi_x$ and $\Pi_z$, together with the expected dynamics {and the imprints expected in the photoelectron momentum distributions}, are provided in Table \ref{tab:OrbitClassification}. 

For some trajectories which are known to not interact strongly with the core, such as orbit 1, the SFA solution can be used as an initial guess and then, by an incremental increase of the Coulomb strength, perturbed towards the genuine semiclassical trajectory. After this, nearby trajectories are found by considering previously solved trajectories on adjacent grid points as initial guesses. There is some subtlety to this step as the exact prescription for which initial guesses should be used to solve which grid points will influence the types of trajectories found. In this case, for a sufficiently small ring around $\textbf{p}_f = 0$, no caustics, which would halt the progress of the solver, are encountered. 

The requirement of such a prescription reduces the versatility of this method. It is necessary to alter it for even the simplest deviations from a monochromatic field, and it will fail when required to find different types of trajectories or for more general field types. For examples of how the standard classification in Table \ref{tab:OrbitClassification} must be altered for bichromatic and elliptically polarized fields, see our previous publications \cite{Rook2022} and \cite{Kim2022}, respectively. Furthermore, in Ref.~\cite{Maxwell2018} it was first evidenced that other distinct orbit types may satisfy the conditions highlighted in the second and third columns of Table \ref{tab:OrbitClassification} and one should not presuppose that the standard behavior will always hold. Nonetheless, we have observed that is not possible for an orbit to fail to satisfy one of the classifications in Table \ref{tab:OrbitClassification} unless it remains bound. Thus, these conditions, although exhaustive, do not provide a unique identification for the trajectories that should be included in the computation of the transition amplitude, and a more flexible approach is called for.
The issue is that the classifications do not always remain true to the spirit of the dynamics highlighted in Ref.~\cite{Yan2010} and used in the standard, boundary version of the CQSFA.

 \section{Forward and hybrid methods}
 \label{sec:novelmethods}
 
Below, we construct two alternative approaches that use the CQSFA formulation allowing for more flexible solutions. The first starts from the CQSFA transition amplitude \eqref{eq:MpPathSaddle}, but uses the sub-barrier part of the CQSFA to build a non-adiabatic ionization rate with Coulomb corrections. Subsequently, we solve the forward problem by launching an ensemble of trajectories, whose initial conditions are sampled from that rate. The second is a hybrid forward-boundary CQSFA method, which starts by launching a large set of Coulomb-distorted trajectories. These trajectories are then used as guesses for the boundary problem instead of relying on pre-existing assumptions about the contributing orbits.

Throughout, unless otherwise stated, we consider a linearly polarized monochromatic field
\begin{equation}
   \mathbf{E}(t)=E_0\sin(\omega t)\hat z. \label{eq:Efield}
\end{equation}
That leads to a vector potential
\begin{equation}\label{eq:field}
\mathbf{A}(t)=A_0\cos(\omega t)\hat z=2\sqrt{U_p}\cos(\omega t)\hat z,
\end{equation}
where $\hat z$ is the polarization direction of the electric field, 
and $U_p$ is the ponderomotive energy. This field choice ensures that the classification in Table \ref{tab:OrbitClassification} holds and facilitates comparison with the standard, boundary-type CQSFA. 

 \subsection{Rate-based forward method} \label{sec:ratebased}

In our rate-based forward approach, we will evaluate the distribution at the detector by using a ``shooting method" \cite{Lein2016,shvetsov2021semiclassical,Li2016PRA} to launch a large ensemble of trajectories. Afterwards, we will bin the final momentum of the trajectories and add coherently the ones lying within the same bin centered at $\mathbf{p}_f$ in momentum space. The ionization probability $\mathcal{P}(\textbf{p}_f)$ then is
\begin{equation}\label{ionization_prob}
\mathcal{P}(\textbf{p}_f)=\biggl|\displaystyle\sum_{j=1}^{n_b} \mathcal{C}(t'_{js})e^{iS_j(\mathbf{\tilde{p}}_s,\textbf{r}_s,t,t'_s)}\biggr|^2.
\end{equation}

The sum is carried over all the trajectories $n_b$ arriving at a given bin. The determinant in the pre-exponential factor in \eqref{eq:MpPathSaddle}, usually called the stability factor, cannot be correctly included in our ``shooting method". As previously pointed out by \cite{shvetsov2021semiclassical,brennecke2020gouy}, the sampling of the trajectories yields a weight that is $1/\det$, instead of the $1/\sqrt{\det}$, predicted here. The term $\mathcal{C}(t'_s)$  will be included subsequently in our calculations when computing PMDs.  However, as our system is ionized from a $1s$ state, we expect it to have a minor influence on the final momentum distribution. This influence will be significant for bound states with angular dependence, but this issue will not be addressed in the present work. For a discussion, see Ref.~\cite{Bray2021}. 

The action in Eq.~\eqref{ionization_prob} can be split into its real and imaginary parts,
 \begin{equation}
S_j(\mathbf{\tilde{p}}_s,\textbf{r}_s,t,t'_s)=\Re S_j(\mathbf{\tilde{p}}_s,\textbf{r}_s,t,t'_s)+\Im S_j(\mathbf{\tilde{p}}_s,\textbf{r}_s,t'_r,t'_s).
 \end{equation}

 The imaginary contribution comes only from the complex integral under the barrier as $S^{\mathrm{prop}}(\mathbf{\tilde{p}}_s,\textbf{r}_s,t,t'_s)$ \eqref{eq:Sprop} is real due to the approximation made upon the tunnel exit. Hence,
 \begin{equation}
     \Im S_j(\mathbf{\tilde{p}}_s,\textbf{r}_s,t'_r,t'_s)=\Im S^{\mathrm{tun}}(\mathbf{\tilde{p}}_s,\textbf{r}_s,t'_r,t'_s),\label{eq:Sim}
 \end{equation}
 where $S^{\mathrm{tun}}$ is given in Eq.~\eqref{eq:Stun} .
 Then, we can write the ionization probability (\ref{ionization_prob}), with exponential accuracy, as
 \begin{equation}\label{ionization_prob1}
\mathcal{P}(\textbf{p}_f)\approx \biggl|\displaystyle\sum_{j=1}^{n_b} \sqrt{W_j(\mathbf{\tilde{p}}_s,\textbf{r}_s,t'_r,t'_s)}e^{i\Re S_j(\mathbf{\tilde{p}}_s,\textbf{r}_s,t,t'_s)}\biggr|^2,
\end{equation}
with
\begin{equation}
    \Re S_j(\mathbf{\tilde{p}}_s,\textbf{r}_s,t,t'_s) =  S^{\mathrm{prop}}+\Re  S^{\mathrm{tun}}\label{eq:Sreal},
\end{equation}
and
\begin{equation}
W_j(\mathbf{\tilde{p}}_s,\textbf{r}_s,t'_r,t'_s)=e^{-2\Im S_j(\mathbf{\tilde{p}}_s,\textbf{r}_s,t'_r,t'_s)},\label{eq:general_ioniz_rate}
\end{equation}
is the instantaneous ionization rate \cite{Ammosov1986,Perelomov1967}, determined by the dynamics under the barrier. 
 
The above equation is similar to those encountered in the implementations of the QTMC \cite{Li2016PRA} and the SCTM \cite{Lein2016,shvetsov2021semiclassical} to find the total ionization probability after binning the final momentum distribution. The main difference between these models and our present implementation will be the derivation of a Coulomb-corrected non-adiabatic ionization rate based on the CQSFA approach, as described in the next session.

\subsubsection{Non-adiabatic ionization rate}
The derivation of the ionization rate for the field, given by Eqs.~\eqref{eq:Efield} and \eqref{eq:field}, will follow the procedure described in Ref. ~\cite{Li2016PRA}. As mentioned before, we will restrict ourselves to a 2-dimensional problem with the initial canonical momentum $\mathbf{p_0}=(p_{0z},p_{0x})$. 
Solving the saddle point equation \eqref{SPEt}, expressing the complex time $t'_s=t'_r+it'_i$, and separating into real and imaginary parts, we obtain the longitudinal canonical momentum at the tunnel exit,
\begin{equation}\label{eq:pz0}
    p_{0z}=-\frac{E_0}{w}\cos(\omega t'_r)\cosh(\omega t'_i),
\end{equation}
and the real part of the tunnel exit \eqref{eq:tunnelExitGeneral} explicitly, 
\begin{equation}\label{eq:z0}
    z_0=\frac{E_0}{w^2} \sin(\omega t'_r)\biggl(1-\sqrt{1+\gamma^2(t'_r,p_{0x}^2})\biggr),
\end{equation}
with the effective Keldysh parameter given by \cite{Li2016PRA}  
$$\gamma(t'_r,p_{0x})=\frac{\omega \sqrt{p_{0x}^2+2I_p}}{|E(t'_r)|},$$
that depends on the amplitude of the electric field at the ionization time and the initial transverse momentum.
Furthermore, we can find the relation between the real and imaginary components of $t'_s$,
\begin{equation}\label{eq:ti}
    \sinh(\omega t'_i)=\gamma(t'_r,p_{0x}).
\end{equation}
We have expressed both the initial longitudinal momentum and the tunnel exit as a function of the transverse momentum and the ionization time; hence only $t'_r$ and $p_{0x}$ remain as independent variables. Next, we will obtain an analytical expression for the instantaneous ionization rate $W(t'_r,p_{0x})$.
Using the above relations and our particular choice of the electric field together with Eq.~\eqref{eq:Stun}, Eq.~\eqref{eq:Sim} gives
\begin{multline}\label{eq:deriveImSj}
    \Im S(t'_r,p_{0x}) =
    \frac{E_0^2}{2\omega^3}\biggl[\biggl(\cos^2(\omega t'_r)+\gamma^2(t'_r,p_{0x})+\frac{1}{2}\biggr)\times\\
    \sinh^{-1}\gamma(t'_r,p_{0x})
     -\frac{\gamma(t'_r,p_{0x})}{2}\biggl(2\cos^2(\omega t'_r) +1\biggr)\\
     \times\sqrt{1+\gamma(t'_r,p_{0x})^2}\biggr]
     -\displaystyle\int_{t_s'}^{t_r'}\Im V(\textbf{r}_0(\tau)d\tau.
\end{multline}

 To derive an analytical expression for 
   \begin{equation}\label{eq:ImV0}
       I_{V_0}= \displaystyle\int_{t_s'}^{t_r'}V(\textbf{r}_0(\tau))d\tau,
   \end{equation}
  we use the long-wavelength approximation, as in \cite{Maxwell2017} to expand the tunneling trajectory $\textbf{r}_0$ \eqref{eq:rTun} around the imaginary component, so that 
\begin{equation}\label{eq:r0longwave}
   \textbf{r}_0(\tau)=(\tau_i-t_i')\biggl[i[\textbf{p}_0+\textbf{A}(t'_r)]-\dfrac{1}{2}\dot{\textbf{A}}(t_r')(\tau_i+t_i')\biggr],
   \end{equation}
 where $\tau_i=\mathrm{Im}[\tau]$. Finally, we obtain the regularized expression as computed in \cite{Maxwell2017a}, 
\begin{multline}\label{eq:V0}
    I_{V_0}(t'_r,p_{0x})= i/\sqrt{-p_{0x}^2+\chi^2}\times\\\ln\biggl[\dfrac{2t'_i(\chi^2-p_{0x}^2)}{\chi\eta-p_{0x}^2+\sqrt{\eta^2-p_{0x}^2}\sqrt{\chi^2-p_{0x}^2}}\biggr]
\end{multline}
with
\begin{align}
     \eta&= i(p_{0z}+A(t_r'))-\dfrac{1}{2}t'_i\dot A(t_r'),\\
     \chi&=i(p_{0z}+A(t_r'))-t'_i\dot {A}(t_r'),
\end{align}
with $p_{0z}$ and $t'_i$ given by Eqs.~ \eqref{eq:pz0} and \eqref{eq:ti}.

Combining together Eq.~\eqref{eq:general_ioniz_rate} and Eq.~\eqref{eq:deriveImSj} we can express the Coulomb corrected non-adiabatic ionization rate $W_C(t'_r,p_{0x})$ as,
\begin{equation}\label{eq:ionization_rate_new}
      W_C(t'_r,p_{0x})=e^{-2\Im S(t'_r,p_{0x})}=W_{V_0}(t'_r,p_{0x})W_0(t'_r,p_{0x}),
\end{equation}
   where $W_{V_0}(t'_r,p_{0x})$ is the Coulomb contribution to the rate given by,
   \begin{multline}\label{eq:expImV0}
    \hspace{-0.5cm} W_{V_0}(t'_r,p_{0x})= e^{-2\Im I_{V_0}(t'_r,p_{0x})}=\\
   \biggl|\biggl[ \dfrac{2t_i'(\chi^2-p_{0x}^2)}{\chi\eta-p_{0x}^2+\sqrt{\eta^2-p_{0x}^2}\sqrt{\chi^2-p_{0x}^2}}\biggr]^{\dfrac{1}{\sqrt{-p_{0x}^2+\chi^2}}}\biggr|^2,
  \end{multline}
 
and $W_0(t'_r,p_{0x})$ contains the rest of the contribution from Eq.~\eqref{eq:deriveImSj}
      
\begin{widetext}
 \begin{equation}\label{eq:W0}
W_0(t'_r,p_{0x})=\exp\biggl[ -\frac{E_0^2}{\omega^3}\biggl(\cos^2(\omega t'_r)+\gamma^2(t'_r,p_{0x})+\frac{1}{2}\biggr)\sinh^{-1} \gamma(t'_r,p_{0x}) -\dfrac{1}{2}\gamma(t'_r,p_{0x})[2\cos^2(\omega t'_r) +1]\sqrt{1+\gamma(t'_r,p_{0x})^2}\biggr]
   \end{equation}
\end{widetext}
\noindent is the non-adiabatic rate without Coulomb correction, only different from the one derived in \cite{Li2016PRA} because of the particular choice of the electric field.
   
Furthermore, the real part of the action along the first part of the contour reads
\begin{multline}\label{eq:real action_tun}
    \Re S^{tun}=\frac{E_0}{\omega^2}p_{0z}\sin(\omega t'_r)(\cosh(\omega t'_i)-1)\\
    +\frac{E_0^2}{8\omega^3}\sin(2\omega t'_r)(\cosh(2\omega t'_i)-1)-\Re I_{V_0},
\end{multline}
 with $I_{V_0}$ given by Eq.~\eqref{eq:V0} in the long wave limit. This will be added to the continuum action according to Eq.~\eqref{eq:Sreal}. 
Our method is similar to that presented in \cite{Li2016PRA,Lein2016,shvetsov2021semiclassical}. Our main contribution is the Coulomb sub-barrier correction included in the ionization rate, based on the CQSFA transition amplitude. We have also added the additional phase coming from the real part of the integral under the barrier \eqref{eq:real action_tun}. This contribution from the Coulomb sub-barrier integral has been shown before \cite{Yan2012} as responsible for shifting the holographic fringes' positions, leading to a better agreement with the TDSE calculations. Studies of such contributions and analytical approximations are also given in our previous paper \cite{Maxwell2017a}.

In our rate-based method, we use accept-reject sampling to obtain the initial transverse momentum $p_{0x}$ and ionization time $t'_r$ from the distribution $\sqrt{W_C(t'_r,p_{0x})}$ \eqref{eq:ionization_rate_new}. By doing this, we require fewer trajectories to resolve the interference patterns, as we sample from the regions where the ionization probability is relevant. Furthermore, it allows us to evaluate the photoelectron momentum distributions at the detector simply by adding the contribution of the phase in the continuum within the different bins in momentum space as,
 \begin{equation}\label{eq:ionizationprop_importantsampling}
\mathcal{P}(\textbf{p}_f)\approx \biggl|\displaystyle\sum_{j=1}^{n_b} e^{i\Re S_j(\mathbf{\tilde{p}}_s,\textbf{r}_s,t,t'_s)}\biggr|^2.
\end{equation}

To avoid contributions from ATI rings we restrict the sampling of the ionization time to one field cycle. Once we have the initial transverse momentum and ionization times, we obtain the longitudinal momentum and tunnel exit positions from Eq.~\eqref{eq:pz0} and Eq.~\eqref{eq:z0}. This gives the complete set of initial conditions to propagate the classical trajectories in the continuum defined by equations \eqref{SPEp} and \eqref{SPEr}. To integrate the equation of motion numerically, we use an adaptive step-size fourth-order Runge-Kutta integrator \cite{Hairer}.

{We use an eight-cycle trapezoidal pulse with two ramp-on, four constant amplitude, and two ramp-off cycles.} Subsequently, the electron moves only under the influence of the Coulomb potential. If, after the pulse is switched off, the electron energy is negative, that is, $E<0$, we interpret this as it is captured in a Rydberg state. Therefore, it will not contribute to the PMD. If, on the other hand, $E>0$ at the end of the pulse, the electron is freed and will reach the continuum. 

The asymptotic momentum $\textbf{p}_a$ of free electrons is calculated from the momentum $\textbf{p}(T_f)$ and position $\textbf{r}(T_f)$ at the end of the laser pulse $(t=T_f)$ using Kepler's solutions \cite{shvetsov2012ionization},
\begin{equation}\label{eq:keppler_sol}
    \textbf{p}_a=p\dfrac{p(\textbf{L}\times \textbf{a})-\textbf{a}}{1+k^2L^2},
\end{equation}
where $\textbf{L}=\textbf{r}(T_f)\times \textbf{p}(T_f)$, and $\textbf{a}=\textbf{p}(T_f)\times \textbf{L}-\textbf{r}(T_f)/r(T_f)$.
Then, we evaluate the photoelectron momentum distribution by binning the final asymptotic momentum of the ionized electrons and adding them coherently within each bin using Eq.~\eqref{eq:ionizationprop_importantsampling}.

\subsubsection{Analysis of the Coulomb sub-barrier correction}
\label{sec:subbarrier}

\begin{figure}[htb]
\centering
\includegraphics[width=1.0\linewidth]{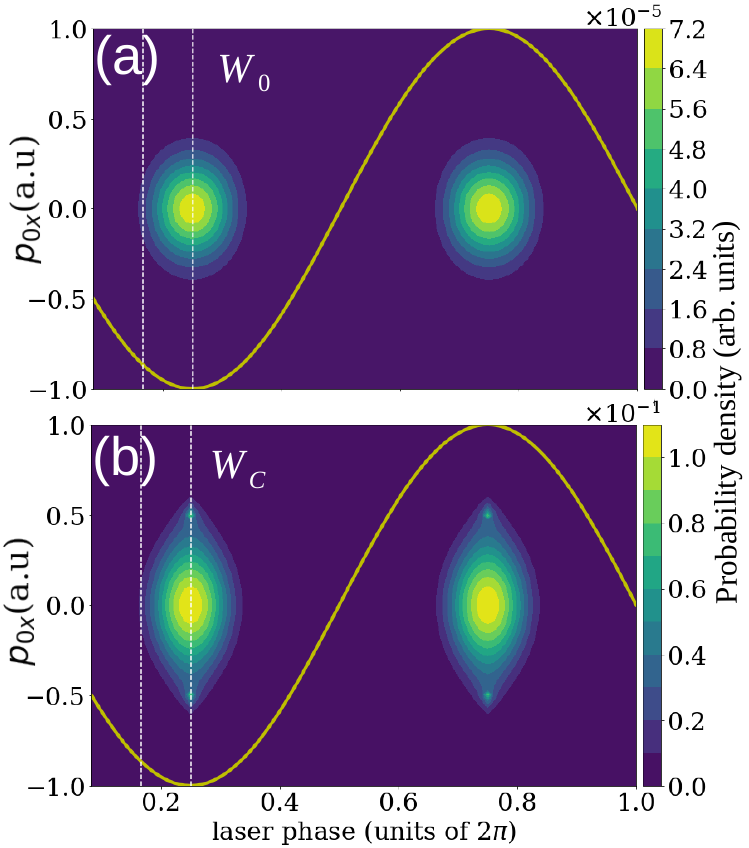}\\
\caption{Transverse momentum distribution at the tunnel exit as a function of the laser phase $\omega t'_r/(2\pi)$ from  $W_0(t'_r,p_{0x})$ (upper panel) and $W_C(t'_r,p_{0x})$ (lower panel). For these plots, we use a hydrogen atom ($I_p=0.5 \; \rm a.u.$) in a laser field of intensity $I=1.5\times 10^{14}\rm \; W/cm^2$ and  wavelength $\lambda=800 \; \rm nm $. The thick solid yellow (light gray) line indicates the laser-field amplitude (up to a scaling factor), and the dashed white lines were drawn at phases $\omega t'_r=\pi/3$ and $\omega t'_r=\pi/2$. The distributions $W_0$ and $W_C$ are computed in arbitrary units. \label{fig:densityplot_W0_W+}}
\end {figure}

In this section, we study how the initial momentum distribution changes due to the influence of the Coulomb sub-barrier correction included in the rate. Figure \ref{fig:densityplot_W0_W+} shows the density plot of the ionization rates, $W_0$ and $W_C$, as a function of the initial transverse momentum $p_{0x}$ and the laser phase $\omega t'_r$. The latter can be mapped into the initial longitudinal momentum $p_{0z}$ using Eq.~\eqref{eq:pz0}, so that broader (narrower) ranges of laser phases correspond to broader (narrower) ranges of $p_{0z}$.

After including the Coulomb potential, we observe an overall increase in the distribution of several orders of magnitude. This is expected from previous work on Coulomb corrected rates \cite{Perelomov1967} and Coulomb distorted approaches such as the Coulomb corrected SFA \cite{Yan2010}, the CQSFA \cite{Lai2015a} and the Volkov-eikonal approximation \cite{Smirnova2006,Smirnova2008}; for a review see \cite{Popruzhenko2014a}. Furthermore, there is a significant broadening of the transverse momentum distribution, and a narrowing of the range of ionization phases $\omega t'_r$ around the times $\omega t'_r= (2n+1)\pi/2=(2n+1)\omega T/4$, where $T$ is the field cycle.  These times correspond to the maxima and minima of the field, and, for the monochromatic wave in this work, also give $A(t'_r)=0$.  
 
 For the parameters in Fig.~\ref{fig:densityplot_W0_W+}, we quantify in Table \ref{tab:widthCoulomb} the variance of the initial momenta, tunnel exit, and tunnel exit position average. The table shows that the Coulomb corrections cause an increase (decrease) in the transverse (longitudinal)  momentum width, while the widths associated with the tunnel exit decrease much less.  The average tunnel exit also decreases if the Coulomb correction is incorporated, which is a consequence of it restricting $\omega t'_r$ to ranges associated with larger field amplitudes. 
 \begin{table}
\centering
    \begin{tabular}{c c c c c}
    \hline
       Rate & $\sigma_{p0x}$ & $\sigma_{p0z} $ & $\sigma_{z_0} $& $\langle z_0\rangle$\\
       \hline
      $W_C$ & 0.28 & 0.45 & 0.59 & 7.16\\
        \hline
     $W_0 $& 0.25 & 0.52 &0.60  & 7.22\\
     \hline
    \end{tabular} 
    \caption{Width of the initial momentum and tunnel exit distributions (width and average) obtained from $W_C$ and $W_0$, using the parameters in Fig.~\ref{fig:densityplot_W0_W+}.}
    \label{tab:widthCoulomb}
\end{table}

Intuitively, the changes introduced by the sub-barrier Coulomb correction shown in Fig.~\ref{fig:densityplot_W0_W+} may be understood as follows.  The Coulomb potential is pulling the electron back, so it needs more energy to overcome the barrier. Therefore, a larger field amplitude will be necessary for it to escape than if the corrections were absent. This will narrow the ranges of $\omega t'_r$ around the peak-field times.  As the parallel momentum is close to zero for electrons freed close to the peak of the field, this translates into higher transverse momentum. Similar Coulomb shifts to the momentum distribution demonstrated before \cite{Torlina2014} have been described as wavepacket deceleration.

In terms of the orbit classification provided in Sec.~\ref{sec:orbitclass}, we expect that the orbits most influenced by the aforementioned broadening will be orbits 1 and 2. An electron along orbit 1 is expected to reach the detector without further interaction or deflection, so that there will be a minimal escape velocity \cite{Lai2015a,Maxwell2017}. This velocity can be achieved either by increasing $p_{0x}$ or $p_{0z}$. As orbit 2 is expected to be a field-dressed hyperbola along which the electron will be deflected but will not undergo hard scattering, the initial transverse momentum must be non-vanishing and relatively large in order for the electron to escape. In this specific regime, the Coulomb potential will contribute to the electron's escape \cite{Lai2015a,Maxwell2017}, which is reflected in an enhancement in the ionization probability. Orbits 3 and 4 have more restricted initial conditions and are expected to behave like hybrid (orbit 3) or rescattered (orbit 4) orbits. Thus, the Coulomb potential will have a stronger influence in the continuum propagation than in the sub-barrier dynamics. This issue will be discussed in more detail in Sec.~\ref{sec:mapping}.
  
\begin{figure}[htb]
\includegraphics[width=1.0\linewidth]{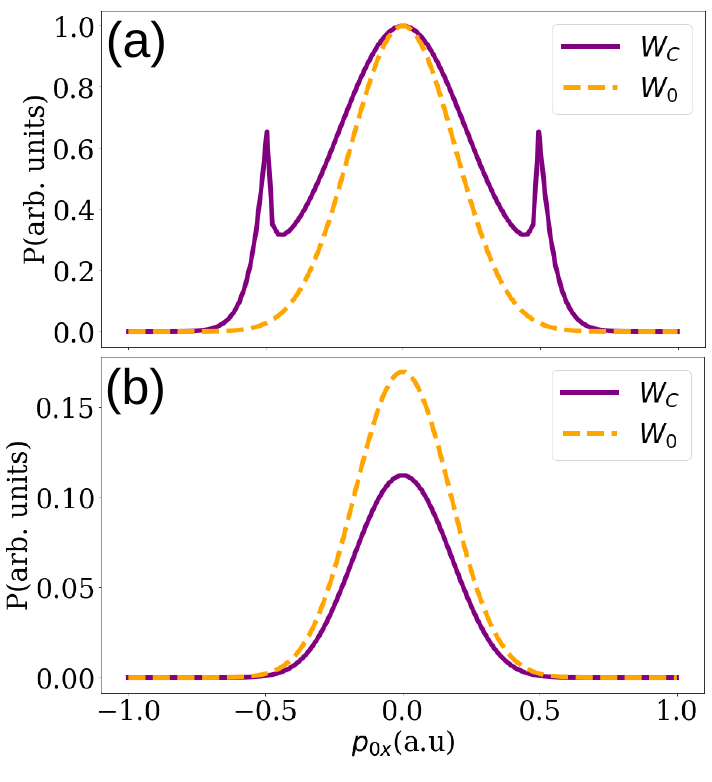}\\
\caption{Instantaneous transverse momentum distributions at the tunnel exit for $\omega t'_r=\pi/2$ (upper panel) and $\omega t'_r=\pi/3 \rm \,a.u$ (lower panel) using the same field parameters as in Figure \ref{fig:densityplot_W0_W+}. The upper panel corresponds to a peak-field time, while for the lower panel, we have taken an ionization phase so that the distribution $W_0(t'_r,p_{0x})$ is touched tangentially near its width (see the dashed lines in  Fig.~\ref{fig:densityplot_W0_W+} for details).  For the dashed orange (light gray) lines, we have employed  $W_0(t'_r,p_{0x})$ while for the solid purple (dark gray) lines we have used $W_C(t'_r,p_{0x})$. For a better comparison, we have normalized the distributions so that they will have the same peak value at the times for which the field amplitude is maximal (upper panel).\label{fig:p0x_fixt}}
\end {figure}

The distribution $W_C( t'_r,p_{0x})$ also exhibits two distinct peaks around $p_{0x}\approx\pm 0.5$ along the peak-field times $\omega t'_r= (2n+1)\pi/2=(2n+1)\omega T/4$, which will be investigated in more detail in Fig.~\ref{fig:p0x_fixt}. In the upper panel of the figure, we plot $W_0(t'_r,p_{0x})$ and $W_C(t'_r,p_{0x})$ for $\omega t'_r=\pi/2$. 

The Coulomb-corrected distribution exhibits two additional peaks, which come from the branch points of the function $1/\sqrt{\mathbf{r}_0^2}$, and the branch cuts originated from them. The components of the tunnelling trajectory $\mathbf{r}_0$ are
\begin{align}
    r_{0z}&=(\tau_i-t_i')\biggl[i[p_{0z}+A(t_r')]-\dfrac{1}{2}\dot A(t_r')(\tau_i+t_i')\biggr]\\
    r_{0x}&=i(\tau_i-t_i')p_{0x},\label{eq:components r0}
\end{align}
so that
\begin{multline}\label{eq:r02expand}
     \mathbf{r}_0^2=\Re(r_{0z})^2+\Re(r_{0x})^2-(\Im(r_{0z})^2+\Im(r_{0x})^2)\\
     +2i\Re(r_{0x})\Im(r_{0x})+2i\Re(r_{0z})\Im(r_{0z}).
\end{multline}

From Eq.~\eqref{eq:components r0} we know that $r_{0x}$ is always imaginary. Therefore, the second and fourth terms in Eq.~\eqref{eq:r02expand} vanish. The branch cuts are located at the points where $\Im \mathbf{r}_0^2=0$ and $\Re \mathbf{r}_0^2<0$. From \eqref{eq:r02expand} we see that this is satisfied when $p_{0z}=0$, and $\omega t'_r=\pi/2 \mod \pi$. 
These peaks are absent in the lower panel of Fig.~\ref{fig:p0x_fixt}, for which the ionization phase $\omega t'_r=\pi/3$ is far from the field extrema. These sub-barrier branching points are of minor relevance for the present paper, as they occur in an initial momentum range far from that of interest. For discussions of branch cuts see \cite{Popruzhenko2014b,Keil2016,Pisanty2016,Pisanty2016a,Maxwell2018b}. 

 \subsection{Hybrid forward-boundary CQSFA}
 \label{sec:HFCQSFA}
 
Next, we will briefly discuss how the CQSFA has been modified in order to be made less reliant on pre-supposing the orbits' dynamics. The CQSFA is formulated and solved as a boundary value problem {in which the final momentum is matched to its initial counterpart at the tunnel exit.} The aim is to find sets of semiclassical trajectories whose final momenta coincide precisely with the grid points of a grid of final momenta. Prior to solving this problem, one needs appropriate guesses for the orbits.  In general, to construct a procedure for finding trajectories, one requires a detailed understanding of the initial to final momentum map. This can be achieved either by using physical intuition and some pre-knowledge of the dynamics or by direct forward propagation of a great number of trajectories. However, the forward propagation itself provides a vast set of initial guesses because the trajectories which have $\textbf{p}_f$ close to a grid point can be readily used as starting points. This approach leads to a significantly more robust method, which is not influenced by one's preconceptions about the types of relevant trajectories. {Computing the relevant orbits in the H-CQSFA consists of three main steps: (i) solving the forward problem, similarly to what is done in the rate-based method described above; (ii) discarding duplicate solutions; (iii) using the remaining final momenta as guesses to solve a boundary problem, in which the initial momenta are refined. } {This binning process was originally developed in \cite{carlsen_advanced_2023}. However, this work focused on short pulses and Maslov phase, and used an adaptive clustering method to find the initial conditions.}

 The first step of the hybrid forward-boundary CQSFA (H-CQSFA) is to numerically calculate the classical trajectories for a large range of initial conditions. For each value of initial momentum, there will be a range of associated saddle-point solutions, which were derived from \eqref{SPEt} and correspond to the various ionization times in the semiclassical picture. For each initial momenta, ionization time pair, the tunnel exit can be evaluated from \eqref{eq:z0}. As in the previous model, the triple $(\mathbf{p}_0,z_0,t')$ suffices as initial conditions for classical continuum trajectories defined by the saddle-point equations \eqref{SPEp} and \eqref{SPEr} to be solved numerically for the duration of the field, which is taken to span four cycles up to the time $T_f$. In order to avoid ATI rings, which may mask the holographic patterns, we consider ionization times $t'$ within a single field cycle. Following the pulse, the final momenta are calculated analytically as the asymptotic momenta of the relevant Kepler hyperbola \eqref{eq:keppler_sol}. 

\begin{figure}
    \centering
   \includegraphics[width=\columnwidth]{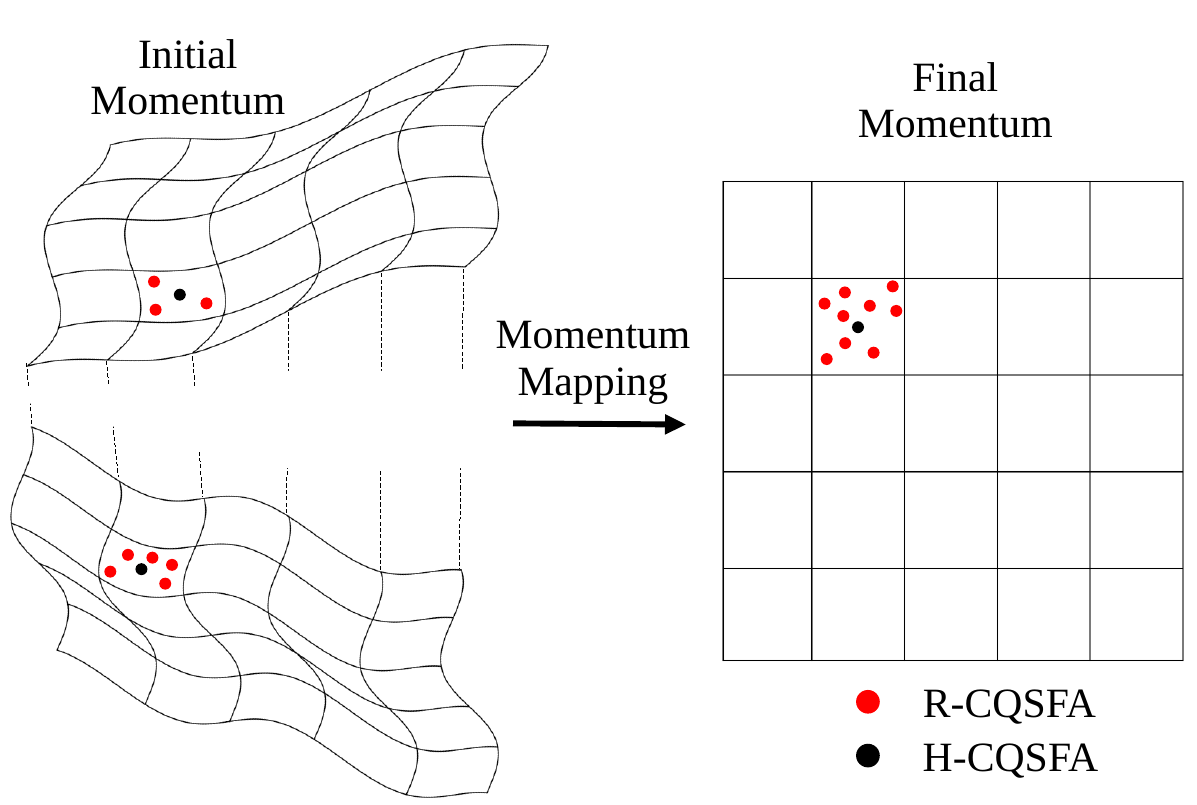}
   \caption{{Schematic representation of the binning process employed in the R-CQSFA (red dots) and the H-CQSFA (black dots). The left side of the figure shows the lines in initial momentum which are mapped to grid lines in final momentum. This is shown here for two distinct regions of initial momentum. In the R-CQSFA, all of the red points correspond to trajectories which are mapped into a single grid square. The amplitude at the centre of the grid square comes from summing over the amplitudes for all of these red points. In the H-CQSFA, the red points would be used as initial guesses for the boundary value problem. The initial guesses are then all refined to the same point at the centre of the grid square. The points which are refined are chosen to avoid having multiple refinements lead to the same initial momentum, but in case they do, duplicates are removed. Therefore, in the H-CQSFA, the amplitude at the centre of the grid square is calculated by summing over only the amplitudes of trajectories which finish exactly at the centre of the grid square.}}
    \label{fig:bin}
\end{figure}

Each grid point of the final momenta can be associated with a bin that has the grid point at its midpoint. The trajectories whose final momenta fall into each bin are then processed. Fig.~\ref{fig:bin} provides a schematic representation of the binning process, 
{where the squares in the initial and final momentum maps represent bins and the dots the momenta associated with the saddle-point solutions mentioned above. The red dots in the initial momentum grid correspond to the whole trajectory ensemble subset within a bin, which is a common feature of the rate-based CQSFA and H-CQSFA. For the forward method, these trajectories are propagated until the final, asymptotic momentum is reached (see red dots in the final momentum map). In contrast, for the H-CQSFA, }trajectories with initial momenta not sufficiently separated are considered to be duplicates and ignored. { Not being sufficiently separated here implies that if the final momenta of the two orbits are continuously deformed to the same value, the initial momentum would be the same value and, thus, they are duplicates.
In Fig.~\ref{fig:bin}, this means that only the black dot is kept. 
}

Subsequently, the remaining trajectories are used as initial guesses for the trajectories which have asymptotic momentum coinciding exactly with the final momentum grid point. {The boundary procedure employed in the standard CQSFA is then adopted, in which the final momenta are constructed as functions of the initial momenta at the tunnel exit. The initial momenta are then refined to match the continuum propagation. This procedure is described in detail in \cite{Lai2015a}.}

The bins are defined arbitrarily, or rather, they are chosen to cover the range of final momenta we would like to show in the photoelectron momentum distribution. The final momenta from the propagation may then fall into one of these bins. It is then determined, based on the content of the bin, 
whether this trajectory will be ``refined” so that it lands right in the center of the bin and can be used to calculate the transition amplitude at its center. During the refinement process, the initial momenta will not be confined by the sampling, and will therefore go wherever it needs to so that its trajectory lands in the center of the bin.

The action and prefactors from Equation \eqref{eq:Stun}, \eqref{eq:Sprop} and \eqref{eq:MpPathSaddle} are calculated numerically such that their amplitudes can be superposed to calculate the probability at a given final momentum. Unless otherwise stated, we consider an initial Gaussian momentum distribution 
\begin{equation}
    \mathcal{P}(p_{0z},p_{0x})=\dfrac{1}{2\pi\sigma_{p0x}\sigma_{p0z}}\exp{-\dfrac{p_{0z}^2}{2\sigma_{p0z}^2}}\exp{-\dfrac{p_{0x}^2}{2\sigma_{p0x}^2}},
    \label{eq:GaussianD}
\end{equation}
for the H-CQSFA, whose widths $\sigma_{p0x},$ $\sigma_{p0z}$ were chosen to match those of the rate $W_C$ given in Table \ref{tab:widthCoulomb}.

Sampling the initial momentum distribution within this approach is more of a matter of including or excluding specific trajectories, as the imaginary part of the action gives the initial weighting. Nonetheless, care must be taken as, if there is a trajectory which was not sampled, and has an initial momentum which does not lie sufficiently close to any of the sampled momenta, it will not be found by the refinement process. Further analysis of the effect of sampling from different distributions will be provided in Sec.~\ref{sec:sampling} by using initial Gaussian distributions of different widths and the same number of grid points.

 \section{Photoelectron momentum distributions}
 \label{sec:PMDs}
 Next, we focus on the photoelectron momentum distributions computed with the methods in the previous sections. For simplicity, unless necessary, we will use the notations $p_x$ and $p_z$ for the final momentum components $p_{fx}$ and $p_{fz}$, respectively. The number of trajectories launched in the rate-based and hybrid methods will vary between $10^7$ and 2 $\times 10^8$, depending on the features of interest. Fully converged spectra are obtained for 2 $\times 10^8$ trajectories, with only minor changes in the PMDs observed for more than $10^8$ orbits. This is in agreement with the parameter range reported in \cite{shvetsov2021semiclassical}.
 
\begin{figure*}[htb!]
    \centering
     \includegraphics[width=0.8\textwidth]{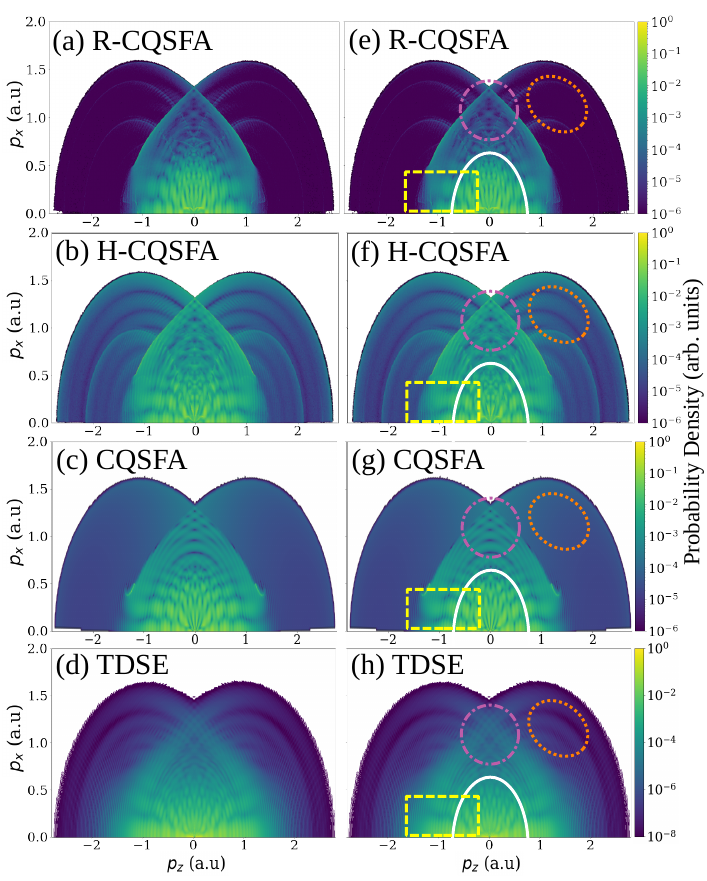}
    \caption{Photoelectron momentum distributions computed for hydrogen ($I_p=0.5 \rm\; a.u$.) in a field of intensity $I=1.5 \times 10^{14}\; \mathrm{W/cm}^2$, wavelength $\lambda=800 \;\rm nm$, {and ponderomotive energy $U_p=0.329 \rm\; a.u.$}, using the rate-based method [panel (a)], the H-CQSFA [panel (b)], the CQSFA [panel (c)] and the Schr\"odinger solver Qprop [panels (d)]. 
    {The panels (e) to (h), on the right column, are replicas of those on the left, with the difference that the holographic features have been highlighted for clarity. The fan is earmarked by a white thin solid line, the spider by a thick dashed yellow rectangle, the carpet and the apex of the caustic by a magenta dotted dashed circle and the annular fringes following the ridges by a dotted oval.}
    For the orbit-based methods, we look at a single cycle incoherently averaging over different unit cells according to \cite{Werby2021,Werby2022}, while for Qprop we use a CEP-averaged one-cycle pulse. In the orbit-based methods, we use a total of $10^8$ orbits for each unit cell. The Qprop outcome was plotted over more orders of magnitude as the rescattering ridges were strongly suppressed for the scale used in the remaining panels. }
    \label{fig:full comparison}
\end{figure*}

In Fig.~\ref{fig:full comparison}, we compare PMDs calculated with the rate-based method [panel (a)], the H-CQSFA [panel (b)], the CQSFA [panel (c)] and the outcome of an \textit{ab-initio} computation [panel (d)], provided by the freely available TDSE solver Qprop \cite{qprop}.  In order to concentrate on the holographic patterns, we consider a single cycle of the field, but add different unit cells incoherently to avoid the arbitrariness associated with the ionization interval having a finite start and endpoint. This is performed by considering different offset phases in Eq.~\eqref{eq:Efield}, that is, setting $\omega t \rightarrow \omega t +\phi$ {therein, with $0\leq \phi \leq 2 \pi$,}  and adding the resulting PMDs incoherently (for details see \cite{Werby2021,Werby2022}). Arbitrary endpoints may lead to asymmetries in the resulting PMDs. For Qprop, we considered a single-cycle pulse but performed an incoherent CEP average in order to eliminate asymmetries.  If more than one cycle is added coherently, prominent above-threshold ionization (ATI) rings stemming from inter-cycle interference are obtained. These rings are well-known in the literature \cite{Becker2018,Faria2020} and have been shown in our previous publications \cite{Lai2017,Maxwell2017,Maxwell2018,Maxwell2018b}. They are not of interest in the present work.  A hybrid coherent-incoherent sum was used in \cite{Rook2022} for monochromatic and bichromatic fields. 

Overall, the agreement of the orbit-based PMDs with the TDSE is reasonably good. All panels in Fig.~\ref{fig:full comparison} show the key holographic structures, such as the fan, the spider and the carpets, a high-energy ridge associated with rescattering and a caustic whose apex is located at the $p_x$ axis. {These features are signposted on the right column [panels (e) to (h)], while on the left [panels (a) to (d)] the full structures can be seen for each method.  } Different coherent superpositions of orbits lead to the distinct holographic patterns. For instance, the interference of orbits 1 and 2 leads to the fan, that of 2 and 3 to the spider and the interference of orbits 3 and 4 to the spiral. These structures have been analyzed in detail in previous publications using the standard, boundary-type CQSFA \cite{Maxwell2017,Maxwell2017a, Maxwell2018,Kang2020,Werby2021,Rook2022}. Therefore, we will keep these discussions as brief as possible in the present article without resorting to specific figures. Sub-barrier Coulomb corrections will also lead to shifts in the interference fringes, but this has been analyzed elsewhere and will detract from the main objective of this work (for discussions, see, e.g., \cite{Yan2012}). 

{The fan and the spider are present throughout, although there are differences in their contrast, and slopes, but the differences are subtle. They are highlighted in the right column by a solid thin circle and a dashed yellow rectangle, respectively.} {The remaining structures exhibit less subtle differences, depending on the method in question.} For instance, the rate-based method and the H-CQSFA [Fig.~\ref{fig:full comparison}(a) and (b)] exhibit secondary ridges at lower energies, and a richer interference structure in the carpet close to the $p_x$ axis, which are absent in the standard CQSFA [Fig.~\ref{fig:full comparison}(c)]. A richer structure for the carpet [see structure highlighted by the dashed-dotted circles in the right column] is also observed in the TDSE computation [Fig.~\ref{fig:full comparison}(d)], although secondary ridges are harder to identify. We also see that the H-CQSFA exhibits additional fringes, which follow the rescattering ridges [see the dotted ovals in the right column]. These fringes are present in the TDSE results, but not in the rate-based method or the standard CQSFA computations. Furthermore, the signal in the outcome of the rate-based method seems to decay much faster away from the $p_z$ axis than for the remaining orbit-based PMDs, which leads to a slightly suppressed interference carpet and a weaker signal close to the rescattering ridges. 

The discrepancies encountered above require a more detailed assessment of several issues pertinent to specific methods, such as how the specific modeling of the sub-barrier dynamics and ionization influence the resulting distributions, how rigorously one may rely upon the orbit classification provided in Table \ref{tab:OrbitClassification}, and how different initial sampling suppresses or enhances particular features.  In the results that follow, we restrict the ionization times to a single field cycle to avoid the presence of ATI rings. This facilitates the study of holographic patterns. For simplicity, we consider a fixed unit cell determined by the electric field \eqref{eq:Efield} {with offset phase $\phi=0$. This 
 means that the time interval used for defining the unit cell will start and end when $E(t)=0$. }
\subsection{Sub-barrier dynamics and single-orbit contributions}
\label{sec:singleorbit}

\begin{figure}[!htb]
\includegraphics[width=1\linewidth]{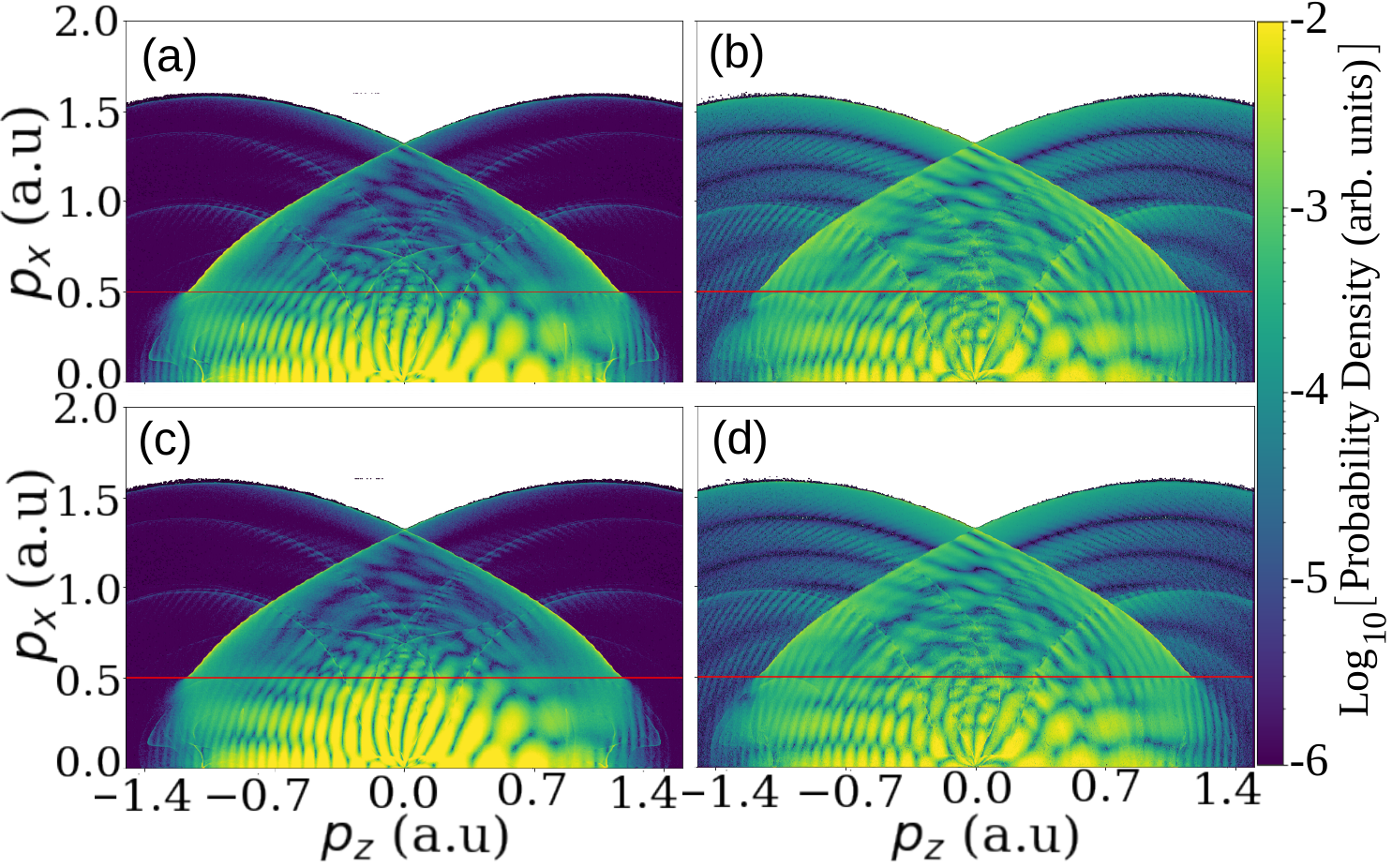}
\caption{Photoelectron momentum distributions (PMDs) computed using ionization times restricted to a single cycle, for the same field and atomic parameters as in Fig.~\ref{fig:full comparison} and a unit cell defined by Eq.~\eqref{eq:Efield}. The left column has been computed with the rate-based forward method using the non-adiabatic ionization rate $W_0$ [panel (a)] and the Coulomb-corrected ionization rate $W_C$ [panel (c)]. The right column has been calculated with the  hybrid forward-boundary CQSFA  removing the Coulomb potential from the tunnelling part of the action [panel (b)] and with full action [panel (d)]. For both methods, we have launched an initial ensemble of $N=2\times 10^8$  trajectories. The solid red (dark gray) lines serve as a guide for the spider and the fan. \label{fig:PMD_compare1}}
\end {figure}

The first question we address in this section is how the results from the forward rate-based method will improve after including the Coulomb correction in the ionization rate, comparing them with the results from the H-CQSFA. These results are presented in Fig.~\ref{fig:PMD_compare1}, for the rate-based method (left column) and the H-CQSFA (right column). For the first one, the presence or absence of sub-barrier Coulomb corrections means that we consider either the ionization rate $W_C$ [Eq.~\eqref{eq:ionization_rate_new}], or the rate $W_0$ [Eq.~\eqref{eq:W0}], respectively. In the H-CQSFA calculations, the Coulomb sub-barrier correction is included by integrating Eq.~\eqref{eq:ImV0} in the complex plane, instead of being incorporated in a rate. Nonetheless, this integral can be switched on and off.  The upper and lower row of Fig.~\ref{fig:PMD_compare1} display the PMDs without and with sub-barrier Coulomb corrections, respectively. 

For the rate-based method, we see how, when using $W_C$ [Fig.~\ref{fig:PMD_compare1}(c)], the PMDs broaden along the perpendicular axis (for comparison, see Fig.~\ref{fig:PMD_compare1}(a), computed with the rate $W_0$). This is expected from the Coulomb correction's effect in the initial transverse momentum distribution (see Fig.~\ref{fig:densityplot_W0_W+}).
The widening along the $p_{x}$ axis happens throughout but mainly affects the fan and the spider. In particular, the fan extends beyond $p_{x}=\pm 0.5 \rm \; a.u$, around the $p_{z}=0$ axis. This happens because the fan results from the interference of direct orbits with field-dressed hyperbolae, and  the spider stems from the interference of field-dressed hyperbolae (orbits 2 and 3), whose initial perpendicular momenta may be large. Therefore, a broader initial transverse momentum will extend the fan to higher momentum regions.
In contrast, the other structures, such as the rescattering ridges, and the carpet-type pattern near the perpendicular momentum axis, result from orbits whose initial momenta are more localized near the polarization axis. This makes them less sensitive to the sub-barrier Coulomb correction. More details will be provided in Sec.~\ref{sec:mapping}, in which the initial-to-final momentum mapping will be analyzed for specific orbits. 

The H-CQSFA outcome, plotted in the right column of Fig.~\ref{fig:PMD_compare1}, shows a similar effect: the Coulomb correction broadens the PMDs in the direction perpendicular to the driving-field polarization, and this broadening is particularly noticeable in the fan [see Fig.~\ref{fig:PMD_compare1}(d)]. The patterns encountered are also similar, but in the rate-based method, the features near the caustic and the carpet near the perpendicular momentum axis are more suppressed. Furthermore, the slope of the spider legs is different. In the hybrid CQSFA approach, they are mostly parallel to the polarization axis, while in the forward method, they bend slightly up. This can be clearly appreciated if we look at how one of the spider legs on the left panel plots cuts the red line, 
 while it seems to cut it tangentially on the right panel plots. Altogether, these results highlight the importance of including the Coulomb potential in the sub-barrier dynamics, either as a correction to rate-based methods or as an extra phase in the semiclassical action for the CQSFA.  Several rescattering ridges are visible in all cases, but the signal is noticeably stronger in the H-CQSFA.

\begin{figure*}[!htb]
    \centering
    \includegraphics[width=1.0\linewidth]{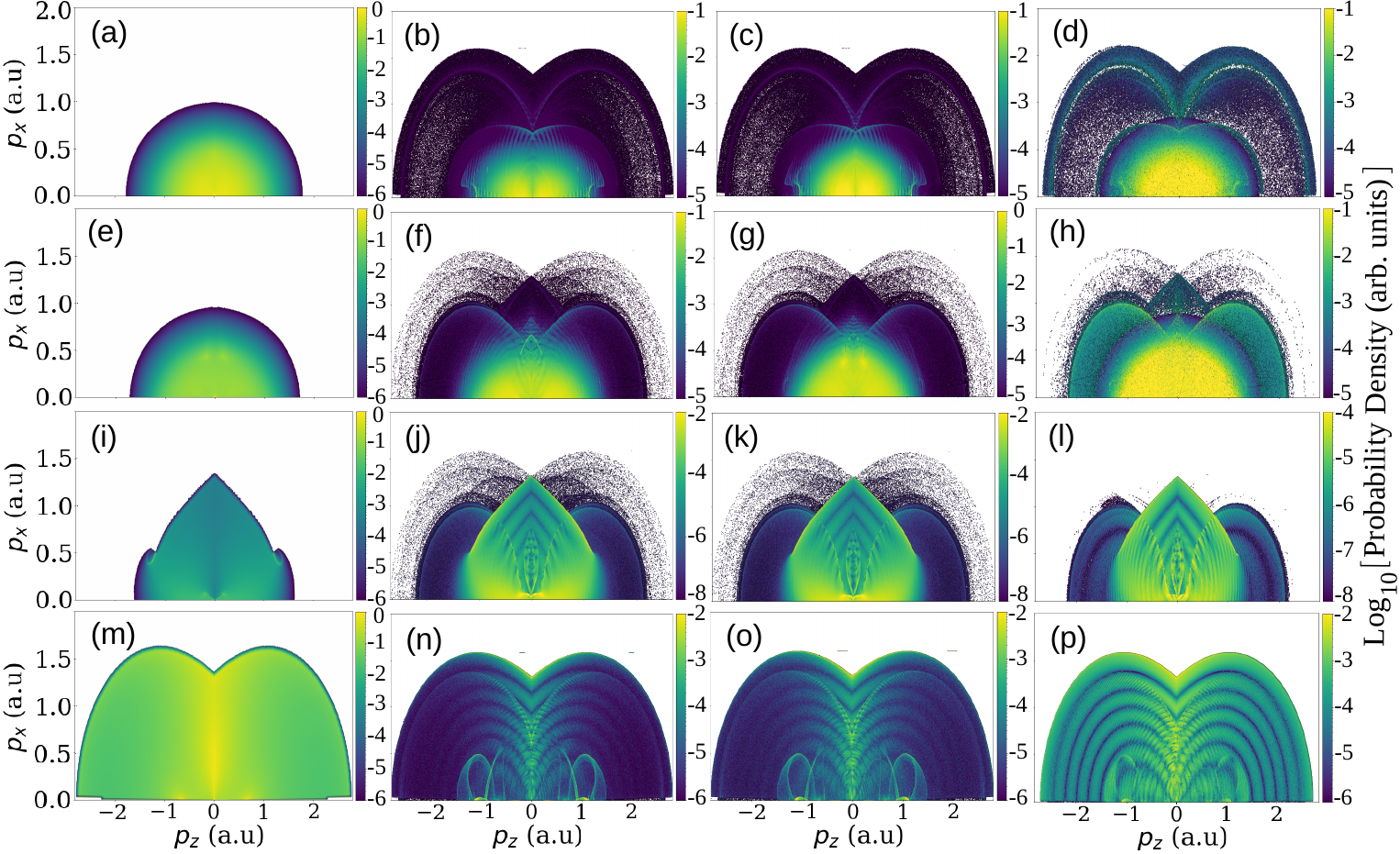}
    \caption{Single-orbit distributions for orbits 1,2,3 and 4, plotted in the first, second, third and fourth row, respectively, following the classification provided in Table \ref{tab:OrbitClassification}.  The distributions were computed for the same parameters as in Fig.~\ref{fig:PMD_compare1} using the the standard pure boundary CQSFA (first column from the left), the forward method with the non-adiabatic ionization rate $W_0$ (second column from the left), the forward method with the non-adiabatic ionization rate with Coulomb correction $W_C$ (third column from the left), and  the hybrid CQSFA with full action (right column). We used $N=2\times 10^8$ initial trajectories in both the rate-based method and the hybrid CQSFA.\label{fig:single_orbits}}
\end{figure*}

Next, we look at single-orbit distributions to investigate how far-reaching the classification introduced in \cite{Yan2010} is in the rate-based or hybrid forward-boundary CQSFA, and also inspect the influence of the sub-barrier corrections. Single-orbit distributions were previously studied within the pure boundary CQSFA approach \cite{Maxwell2017}, and in particular, the effect of the Coulomb integral under the barrier $I_{V_0}$ was addressed in \cite{Maxwell2017a}. These distributions are displayed in Fig.~\ref{fig:single_orbits}, and the orbits are classified according to the conditions in Table \ref{tab:OrbitClassification}.

{The left column of the figure exhibits the standard CQSFA results. The contributions of orbits 1 and 2, displayed in Figs.~\ref{fig:single_orbits}(a) and (e), exhibit the expected imprints (see the fifth column in Table \ref{tab:OrbitClassification}): The PMDs are centered at $(p_z,p_x)=(0,0)$ and decay around the energy of $2U_p$. This is consistent with the expected dynamics of both orbits, which interact weakly with the core and can be associated with direct orbits in the Coulomb-free SFA. The contribution from orbit 3, shown in Fig.~\ref{fig:single_orbits}(i), is bounded by a caustic, which is a qualitatively different behavior from orbits 1 and 2. Finally,  there is a high-energy rescattering ridge in the contributions from orbit 4, as shown in Fig.~\ref{fig:single_orbits}(m). This is expected from the standard definition of this orbit, which is assumed to be rescattered from the start.  Thus, a pure boundary problem with pre-selected behaviors allows us to define the dynamics associated with orbits 1-4 neatly.}

{In contrast, for the rate-based and hybrid CQSFA,} the conditions upon the tunnel exit and the transverse momentum component {given in Table \ref{tab:OrbitClassification}} are insufficient to enforce the dynamics typically associated with orbits 1, 2, 3 and 4. This is evidenced by rescattering ridges and features not necessarily associated with `direct' orbits, such as caustics, present for the contributions of orbits 1 and 2. {These features occur in addition to the expected imprints in Table \ref{tab:OrbitClassification} associated with orbits 1 and 2, which are also present}. Examples are the primary ridge associated with rescattering at energy $10U_p$, which is seen very clearly for orbit 1 [Figs.~\ref{fig:single_orbits}(b) to (d)] and are slightly less intense for orbit 2 [Figs.~\ref{fig:single_orbits}(f) to (h)], as well as secondary ridges at lower photoelectron energies for the distributions stemming from orbit 2  [Figs.~\ref{fig:single_orbits}(f) to (h)].  

{The PMDs resulting from orbit 3 also exhibit, apart from the central caustic common to all methods and signposted in Table \ref{tab:OrbitClassification} as its typical imprint, interference fringes and secondary ridges  [see Figs.~\ref{fig:single_orbits}(j) to (l) compared to Fig.~\ref{fig:single_orbits}(i)]. }
The secondary ridges observed at lower photoelectron energies are associated with longer orbits, which return at least after 1.5 cycles, but, nonetheless, satisfy the conditions imposed upon orbits 2 and 3. This implies that care must be taken with the standard CQSFA orbit classification if forward or hybrid methods are used. 

The number and intensity of those secondary ridges vary according to the method employed, being more prevalent in the rate-based approach. There are up to four secondary ridges for the distributions associated with orbit 2 in both rate-based and hybrid methods [Figs.~\ref{fig:single_orbits}(f) to (h)]. For orbit 3, four secondary ridges are only visible for the rate-based approach [Figs.~\ref{fig:single_orbits}(j) and (k)], while for the hybrid CQSFA, there is only a low-energy ridge   [Fig.~\ref{fig:single_orbits}(l)]. 
For the rate-based method, the primary rescattering ridge is also observed for orbit 3 [see Figs.~\ref{fig:single_orbits}(i) and (j)], while it is absent for the H-CQSFA [see Fig.~\ref{fig:single_orbits}(k)]. Furthermore, the contributions of orbit 2 also exhibit a caustic near the perpendicular momentum axis, which traditionally is associated with orbit 3 \cite{Maxwell2017} [see Figs.~\ref{fig:single_orbits}(f) to (h)].

Another noteworthy feature is that, for the rate-based approaches, the contributions of orbit 3 around the polarization axis are stronger than those of the CQSFA computations. Let us remark that the overall shape of orbit 3 distribution within the CQSFA framework is only obtained when the stability factor $\det[{\partial\mathbf{p}_s(t)}/{\partial \mathbf{p}_s(t'_s)} ]$ given in Eq.~\eqref{eq:MpPathSaddle} is incorporated \cite{Maxwell2017}. As discussed in section \ref{sec:ratebased}, the forward propagation method is implicitly taking this into consideration but with a wrong weight $1/|\det|$ instead of the $1/\sqrt{|\det|}$ employed in both CQSFA calculations, being this the cause of the enhanced contribution around the polarization axis. 

Finally, we observe interference patterns in the single-orbit distribution of orbits 3 and 4 from the rate-based method and hybrid CQSFA. There are fringes following the caustic whose extrema are located near ($p_{z},p_{x})=(0,1.3\; \rm a.u.)$ for orbit 3, clearly visible in panels (j) to (l), and annular fringes following the low-energy ridge in Fig.~\ref{fig:single_orbits}(l). {The interference fringes indicate that more than one type of orbit can be classified as ``orbit 3", and that their contributions to the PMD are being coherently superimposed.}

Further annular interference patterns are also present for orbit 4, following the primary ridge [see Figs.~\ref{fig:single_orbits}(n) to (p)]. For the H-CQSFA, these patterns are equally strong throughout, while for the rate-based method, they are more prominent at the low-energy end of the primary rescattering ridge. {The annular patterns resemble the intra-cycle interference obtained for the rescattered Coulomb-free SFA. In fact, it is well known that rescattered SFA orbits occur in pairs that nearly coalesce close to the maximum photoelectron energy \cite{Faria2002}. This implies there are two types of orbit 4, one of which is being left out by the standard CQSFA. }

Overall, there is a better agreement of the single orbit distributions from the rate-based method with the {remaining CQSFA calculations} when employing the Coulomb-corrected rate. This is more noticeable for the contributions of orbits 1 and 2, which get elongated along the $p_{x}$ axis and narrowed along the polarization axis. For orbit 2 we observe a suppression at $p_{z}=0$ and two bright spots, in the rate-based approach employing $W_C$ and in the boundary CQSFA calculations {[see Figs.~\ref{fig:single_orbits}(e) and (g)]} \cite{Maxwell2017a}. These spots are associated with the branching points that occur for the sub-barrier corrections at the field extrema (see Sec.~\ref{sec:subbarrier}). They are more pronounced for the rate-based approach, as it uses an analytical approximation that overestimates the contribution from the first part of the contour \cite{Maxwell2017a}. The peaks are washed out in the H-CQSFA, as a consequence of having the contribution of several types of orbits falling under the classification of orbit 2.

\subsection{Initial sampling and holographic patterns}
\label{sec:sampling}

One important difference between the rate-based method and the H-CQSFA not addressed so far is related to how the initial conditions are sampled. 
In the rate-based method, the initial transverse momentum and ionization time are sampled either from $W_0$ or $W_C$. This has the advantage that most of the sampled trajectories are located in the higher probability region. For the H-CQSFA, so far, we have considered an initial Gaussian distribution whose parameters mimic those in the rate $W_C$.  

This section aims to assess the impact of sampling from different initial conditions on the different holographic patterns more systematically. In the rate-based method, sampling from an arbitrary distribution will require correcting the weight of the trajectories at the detector, while, in the H-CQSFA framework,  keeping a fixed number of points will imply that we let in or out certain regions in the initial momentum space. 

\begin{figure}[!htb]
  \includegraphics[width=1.0\linewidth]{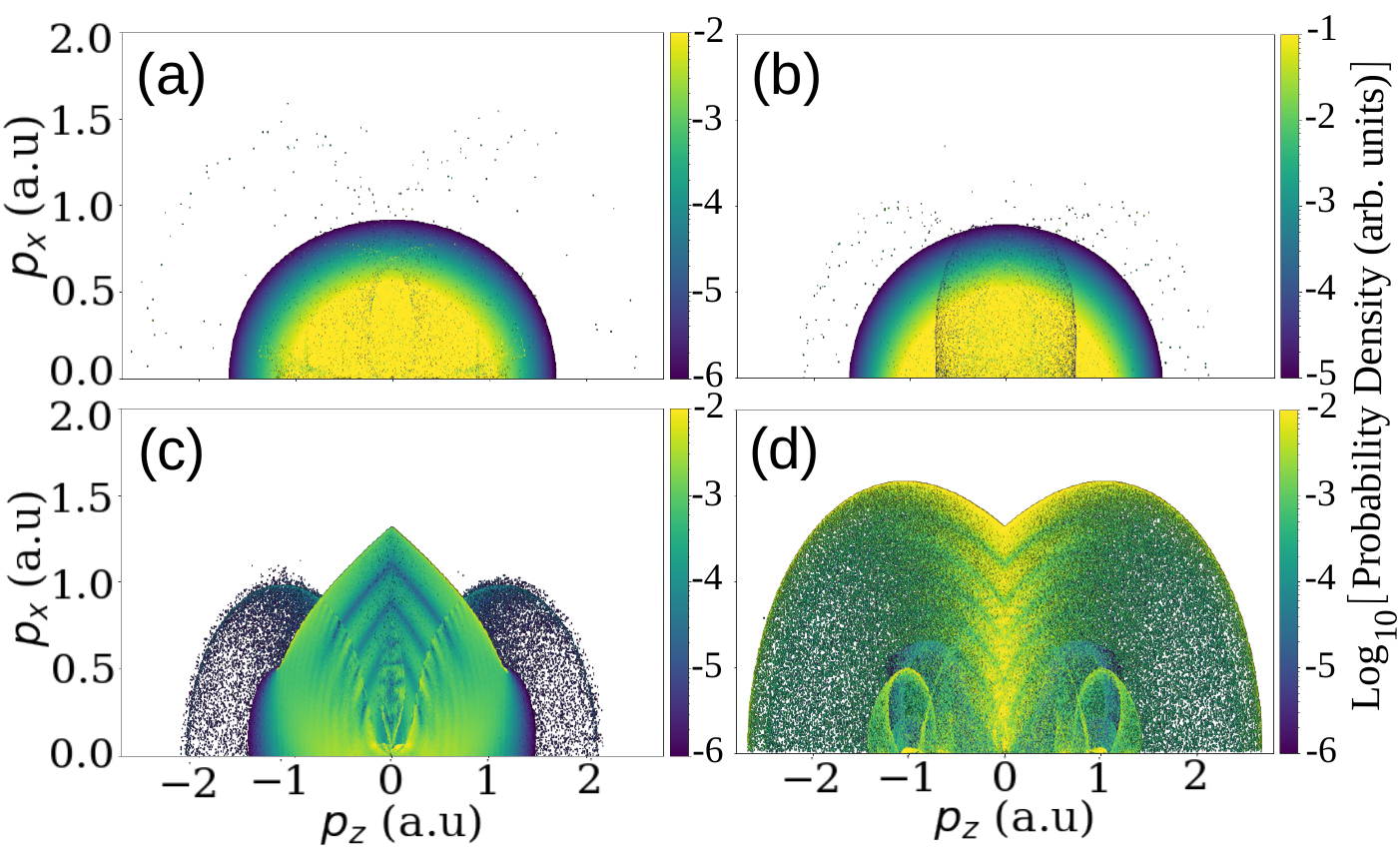}
  \caption{Single-orbit photoelectron momentum distributions calculated with the H-CQSFA for the same parameters as in Fig.~\ref{fig:single_orbits}, but considering a uniform initial sampling of $N=2\times 10^8$ trajectories. Panels (a), (b), (c) and (d) refer to the contributions from orbits 1, 2, 3 and 4, respectively.}\label{fig:single_orbswidth}
\end{figure}

As a first test, in Fig.~\ref{fig:single_orbswidth}, we plot single-orbit distributions computed with the H-CQSFA, in which the initial conditions were sampled from a single uniform grid. The distribution associated with orbits 1 and 2, shown in Figs.~\ref{fig:single_orbswidth}(a) and (b), resembles those obtained with the boundary CQSFA, displayed in Figs.~\ref{fig:single_orbits}(a) and (e). There are no rescattering ridges or caustics, which leads to the conclusion that the orbits causing such features are not being sampled. This is in striking contrast to the results shown in Figs.~\ref{fig:single_orbits}(d) and (h) obtained with an initial Gaussian sampling, which shows these structures. 
The secondary ridge for orbit 3, displayed in Fig.~\ref{fig:single_orbswidth}(c), is also less defined than that obtained with the Gaussian sampling [Fig.~\ref{fig:single_orbits}(l)], and the annular interference structures associated with orbit 4 [see Fig.~\ref{fig:single_orbswidth}(d)] are only present near the perpendicular momentum axis, while for the previous sampling they are visible throughout [see Fig.~\ref{fig:single_orbits}(p)]. This implies that a narrower initial distribution probes the momentum region associated with rescattering in more detail.
This shows that care must be taken when sampling, as the initial momenta lie in different regions and may be densely clustered. Hence, a single uniform sampling distribution may not include all relevant initial momentum points or will inefficiently sample dense clusters. Thus, it is important to investigate alternative sampling distributions such as a Gaussian, or alternatively use adaptive sampling such as in \cite{carlsen_advanced_2023}.

 Next, we will sample the initial transverse and longitudinal momentum from a Gaussian distribution as defined in Eq.~ \ref{eq:GaussianD}, whose widths $\sigma_{pox}$, $\sigma_{poz}$ will be altered to form a narrow $\mathcal{G}_n$, medium $\mathcal{G}_m$ and broader $\mathcal{G}_b$ Gaussian. The standard deviation of the narrow Gaussian along the transverse direction is equal to the one of $W_0$, for $\mathcal{G}_m$ it was chosen close to the standard deviation of $W_0$ along the polarization axis, and for $\mathcal{G}_b$ an arbitrary value above the width predicted by the rates. For simplicity, the width was set equal along each direction. The number of trajectories is kept constant at $10^8$ to illustrate the issue better. 
 The values for $\sigma_{pox}$, $\sigma_{poz}$ are given in Table \ref{tab:width}, for clarity we also added the widths of the distributions $W_0$ and $W_C$.

\begin{table}[!htb]
    \centering
    \begin{tabular}{c c c c }
    \hline
       Distribution & $\sigma_{pox} \rm\; (a.u)$  & $\sigma_{poz}\rm\; (a.u)$  \\
       \hline
       $W_C$ & 0.28 & 0.45 \\
        \hline
       $W_0$ & 0.25 & 0.52 \\
       \hline
        $\mathcal{G}_b$ & 0.75 & 0.75 \\
       \hline
       $\mathcal{G}_m$& 0.5 & 0.5 \\
       \hline
      $\mathcal{G}_n$& 0.25 & 0.25 \\
       \hline
    \end{tabular}
    \caption{Widths $\sigma_{p0x}$ and $\sigma_{p0z}$ of the initial momentum distributions obtained from $W_C$ and $W_0$, respectively, compared with the width of the arbitrary Gaussian distributions.}
    \label{tab:width}
\end{table}

Weighted initial sampling in the H-CQSFA may potentially increase efficiency. However, care must be taken because, depending on the sampling, certain types of trajectories may be included or left out, and this will have an effect on the overall patterns observed. 
On the other hand, in the rate-based method, sampling from an arbitrary distribution will require correcting the weight of the trajectories at the detector when calculating the transition probability combining equations \eqref{ionization_prob1} and \eqref{eq:ionization_rate_new}.

\begin{figure}[!htb]
    \centering
    \includegraphics[width=1\linewidth]{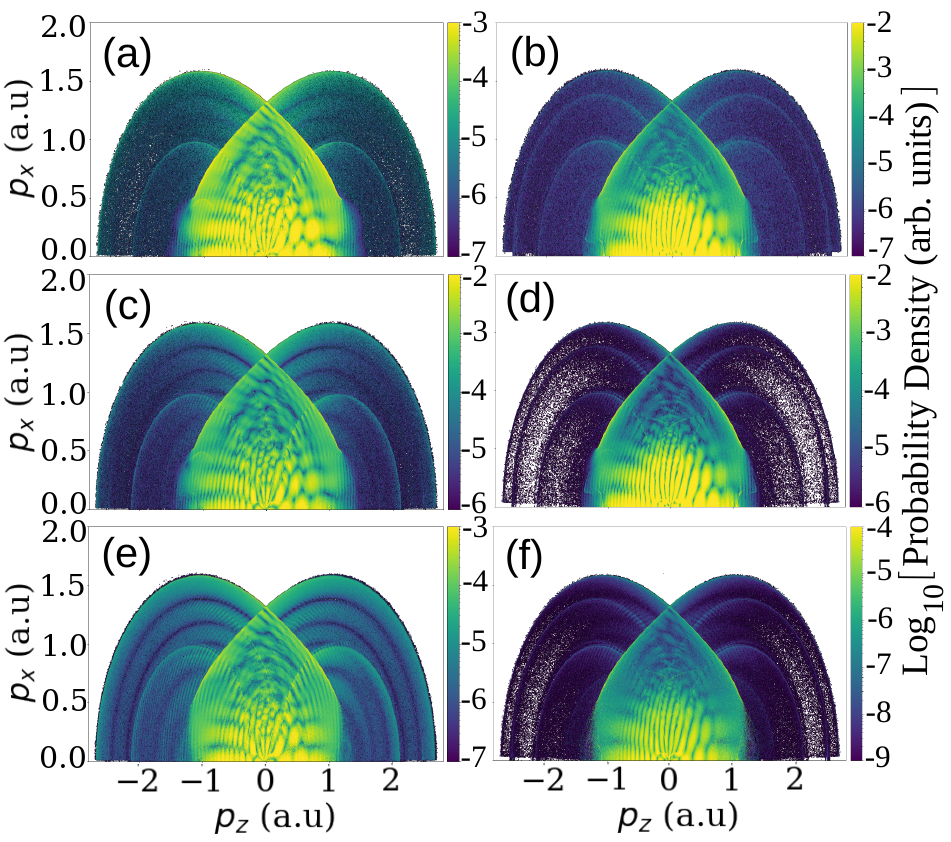}
    \caption{PMDs after sampling from a Gaussian function with different widths $\mathcal{G}_b$ (top row), $\mathcal{G}_m$ (center row), and $\mathcal{G}_n$ (bottom row) as given in Table \ref{tab:width}.  The left panels were computed with the hybrid forward-boundary CQSFA, and the right panels with the forward method, using the same initial Gaussian distributions and correcting the weight at the detector. The figure uses an initial ensemble of $N=10^8$ trajectories.}
    \label{fig:arbitrary_Gaussians}
\end{figure}
In Fig.~\ref{fig:arbitrary_Gaussians} the photoelectron momentum distributions obtained from the different initial distributions are shown in order of decreasing width from  top to bottom. On the left column, we have the results from the H-CQSFA while on the right we have the results from the rate-based method. Fig.~\ref{fig:arbitrary_Gaussians}(a) shows that by using the broad Gaussian $\mathcal{G}_b$  the high-energy ends of the spider, located around parallel momenta $p_{z}=\pm 1 \;\rm a.u.$ and perpendicular momenta $|p_{x}|< 0.5\;\rm a.u$ ., are elongated, compared to the plots with narrower Gaussians $\mathcal{G}_m$ and $\mathcal{G}_n$, for example in Fig.~\ref{fig:arbitrary_Gaussians}(e). Additionally, the fan, which is the interference structure near the threshold $(p_{z},p_{x})=(0,0)$ 
for low parallel momentum and up to around 0.5 in perpendicular momentum, is wider in Fig.~\ref{fig:arbitrary_Gaussians}(a) compared to Fig.~\ref{fig:arbitrary_Gaussians}(c).  As the Gaussian distributions narrow, the spider shortens, losing its high-energy ends, and the fan becomes less spread. In contrast, the rescattering ridges and the interference rings in higher momentum regions become better defined. For instance, Fig.~\ref{fig:arbitrary_Gaussians}(e),  computed for the narrow Gaussian, exhibits truncated spider-like fringes, a somewhat blurred fan, but at least two distinct rescattering ridges, and circular interference fringes following the ridges. There is also a marked improvement in contrast for the carpet-like structure forming near the $p_{x}$ axis and close to the caustic around the $p_x$ axis.  Additional interference structures with the shape of that caustic, which have been identified for the single-orbit distributions for orbit 3, are also visible regardless of the initial sampling taken. 
The reason a more focused sampling around the core leads to these clearer rings is due to better sampling of re-scattering trajectories, which often start close to the core.  This is explored in more detail in the next section. 

These observations compare favorably to the results already presented in Fig.~\ref{fig:PMD_compare1}, where the Coulomb corrected rate-based model is investigated. They concluded that the Coulomb correction during tunneling leads to a broader transverse momentum distribution, which allows a better probing of the fan. An essential difference from the behaviors observed in Fig.~\ref{fig:PMD_compare1} is noticed in the legs of the spider. When using the different ionization rates to sample the initial conditions, the width of the momentum distribution along the polarization axis was in both cases above the width selected for the narrow Gaussian as shown in Table \ref{tab:width}. Therefore, the high-energy ends of the spider were not appreciably affected. Overall the plots on the right column of Fig~\ref{fig:arbitrary_Gaussians} show similar effects to those obtained from the hybrid CQSFA. 
Finally, we must highlight that in the bottom panels, the amplitude of the PMD is significantly lower than in the medium and top panels. 

\section{Momentum mapping}
\label{sec:mapping}

The features discussed in the previous section can be understood in greater depth by looking at the initial to final momentum mapping for specific sets of orbits. Here we will address the question of what momentum ranges $\mathbf{p}_0$ at the tunnel exit lead to specific final momenta $\mathbf{p}$. These studies are necessary as the joint influence of the driving field and the binding potential will modify the electron momenta during the continuum propagation. Moreover, they will shed light on the initial momenta leading to specific holographic features and why a certain sampling highlights particular momentum ranges and structures. 

We will perform this mapping for orbits 1 to 4 according to the classification in Table \ref{tab:OrbitClassification}, and compare our results with the single-orbit distributions in Sec.~\ref{sec:singleorbit}. Because we have observed, using the single-orbit PMDs, that both for the forward and the hybrid methods there are unexpected features associated with rescattered orbits, such as ridges for orbits 1, 2 and 3, we will classify the dynamics further using the tunnel exit $z_0$ and the Bohr radius $r_0$ as parameters. This classification has been first employed in \cite{Maxwell2018} within the context of the boundary CQSFA and uses the distance of closest approach $r_c$ of an electron along a specific orbit. If during the continuum propagation, $r_c$ lies within the region $r_0<r_c<|z_0|$, where $|z_0|$ is the radial distance from the origin determined by the absolute value of the tunnel exit, we assume that the electron has undergone a soft collision. This means that the residual potential was able to deflect the electron and bring it closer to the core than the tunnel exit but has not reached a region for which the potential is dominant. If the distance of closest approach $r_c$ is smaller than the Bohr radius $r_0$, we consider the electron  has a hard collision with the core. Within this classification, a direct orbit implies $r_c>|z_0|$ throughout. { A summary of these filtering criteria is provided in Table \ref{tab:mapping}, together with the features observed for the final momenta. For the initial momenta, we have observed that a wide range of fine structures, such as islands concentrated in narrow momentum ranges, lead to caustics and ridges, but may also be associated with rare events in which there is barely deflection}. Unless otherwise stated, we will focus on a comparison between the rate-based method and the standard CQSFA. The conclusions drawn for the rate-based scenario are also valid for the H-CQSFA. 

{An important difference between the forward and hybrid approaches and the standard CQSFA is that in the former cases a Gaussian initial sample was taken, while, in the CQSFA, a radial grid for the final momentum was used. These choices will reflect themselves in the initial momentum mapping: For the CQSFA, the occupied regions in the momentum plane appear as shaded areas, while, for the remaining methods the density of points is determined by the initial sampling distribution. }
\begin{table}[!htb]
\centering
    \begin{tabular}{l l l }
    \hline
Filtering  & Dynamics  & Final momenta \\ \hline \hline
$r_c>|z_0|$ & Direct &  Spread around the origin    \\ \hline
 $r_0<r_c<|z_0|$ &  Soft scattered &  Caustics \\ \hline
 $r_c<r_0$ & Hard scattered  & Ridges and caustics \\  \hline
\end{tabular}
\caption{Summary of the spatial filtering employed to classify th orbits' dynamics (first and second column), together with the features encountered in the final momentum mapping. \label{tab:mapping} }
\end{table}

 \begin{figure}[!htb]
    \centering
    \includegraphics[width=1\linewidth]{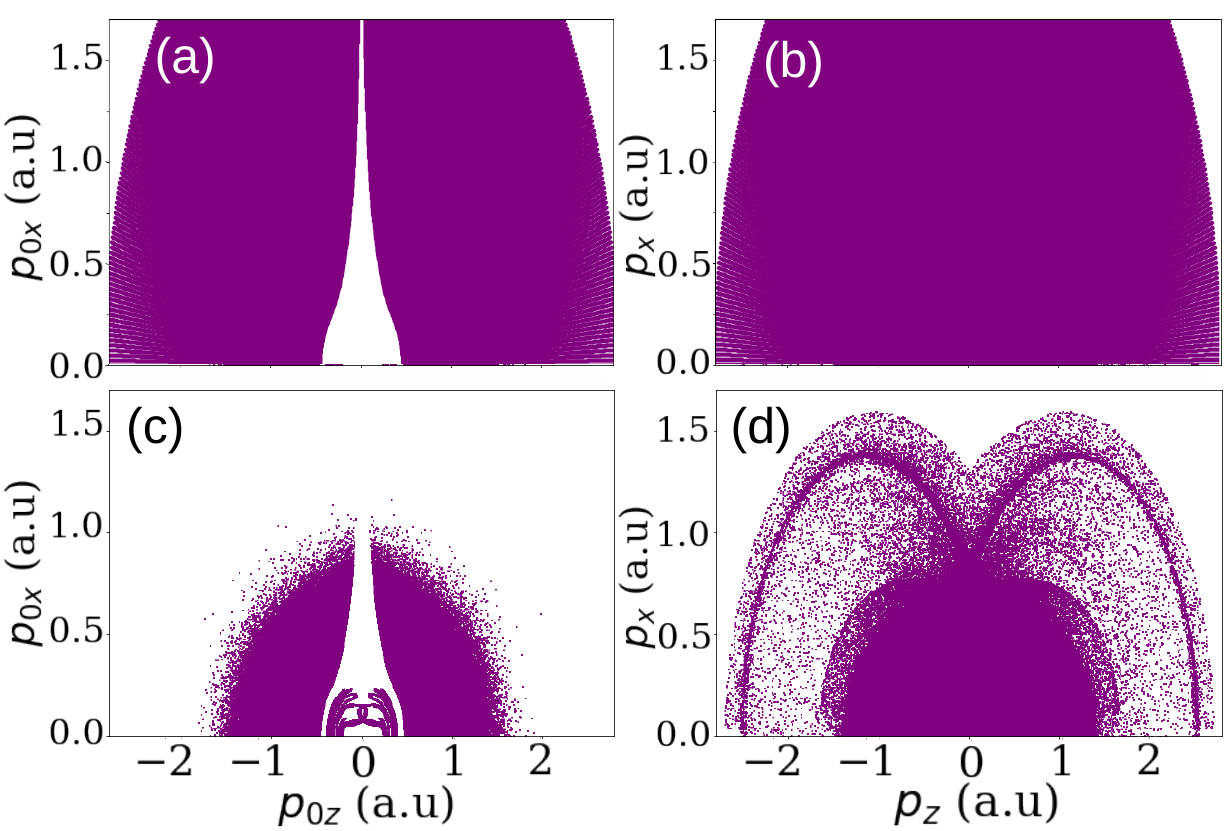}
    \caption{Initial and final momentum of Orbit 1 classified according to Table~\ref{tab:OrbitClassification}, left and right panels, respectively. The first row shows the results from the boundary CQSFA, and the second row the results from the forward rate-based method using $N=10^7$ initial orbits.}
    \label{fig:orbits1}
\end{figure}
\subsection{Orbit 1}

\begin{figure}[!htb]
    \centering
    \includegraphics[width=1\linewidth]{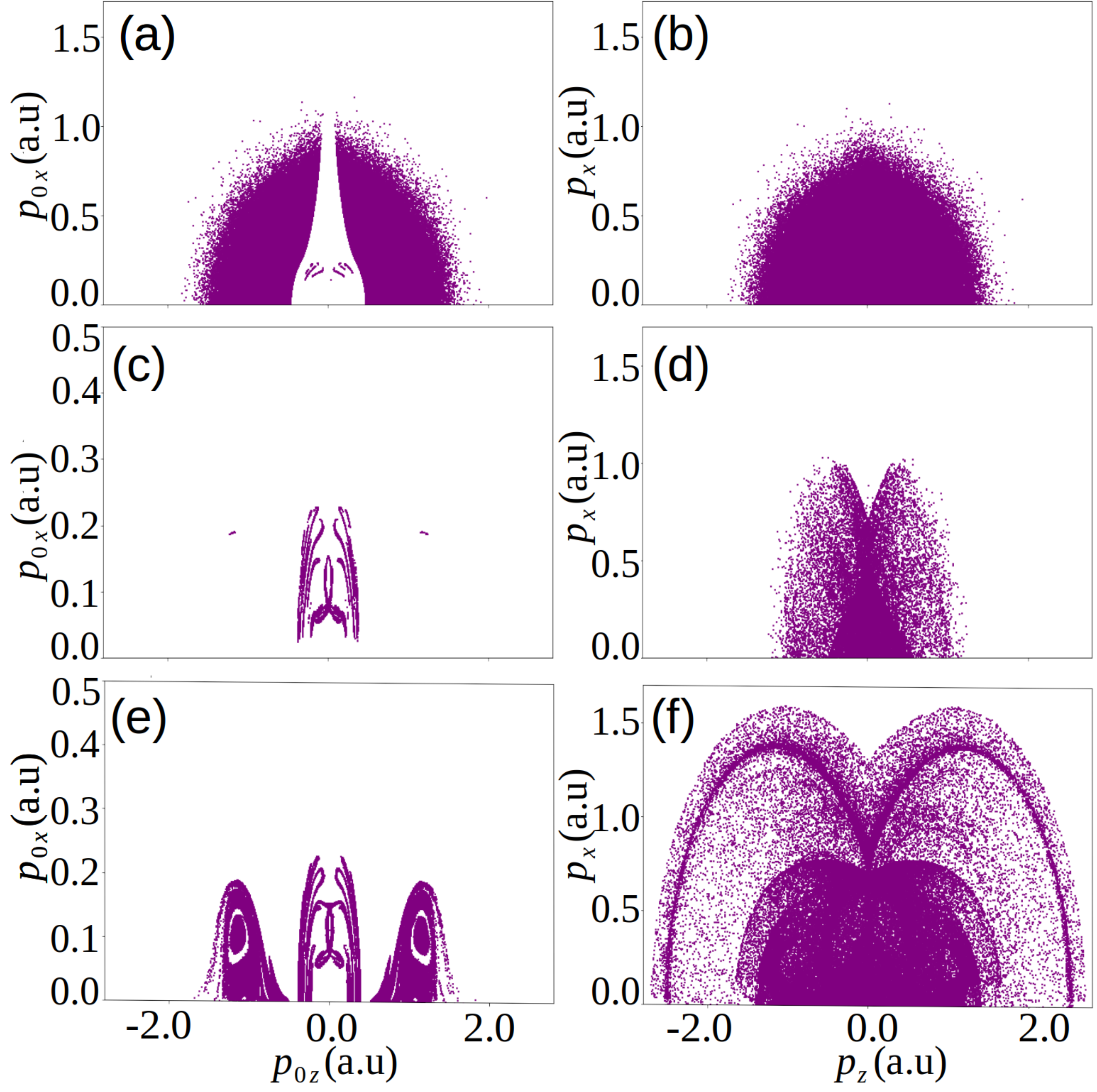}
    \caption{Initial and final momenta of orbit 1, left and right panels, respectively, classified according to the spatial filter using the rate-based method results with $N=10^7$ initial orbits. The top, middle, and bottom panels show the direct, soft-colliding, and hard-colliding orbits, respectively. }
    \label{fig:orbits1filter}
\end{figure}

Fig. \ref{fig:orbits1} shows the initial (panels (a) and (c)) and final (panels (b) and (d)) momenta for orbit 1 following the classification given in Table \ref{tab:OrbitClassification}. The first and the second rows shows the results obtained with the CQSFA and R-CQSFA, respectively. Throughout, the initial momentum distributions exhibit a flame-shaped gap, with the R-CQSFA yielding some structure inside the flame and rescattering ridges that are not present in the boundary CQSFA. 
Further insight is achieved by applying our spatial filter to the results from the forward method, as shown in Fig.~\ref{fig:orbits1filter}. The upper row of the figure shows the orbits that reach the detector without getting closer to the core than the tunnel exit. These are expected to be direct orbits, which escape the Coulomb attraction and reach the detector without further interaction with the target. The initial momenta of these orbits exhibit the flame-shaped gap displayed in Fig.~\ref{fig:orbits1}, which can be understood as follows. Because the Coulomb force pulls the electron back to the core, an electron with vanishing initial momentum would be trapped. Thus, to reach the detector with vanishing momentum $\mathbf{p}=0$, the electron must escape with a non-vanishing momentum $\mathbf{p}_0$.The corresponding momentum distribution at the detector, plotted in Fig.~\ref{fig:orbits1filter}(b), is centered at $(p_{z},p_{x})=(0,0)$ and, as expected, does not exhibit rescattering ridges. The Coulomb attraction has also closed the flame-type gap, as it affects the orbits with low initial velocity the most and those with high initial velocity only marginally. 
Let us also note that there is some structure inside the flame in the initial momentum distribution of the direct orbits obtained with the forward method. These are rare events, and a closer look at the dynamics of these orbits has revealed that they are slightly deflected by the core. However, the distance of closest approach is always above the tunnel exit and they do not lead to rescattering ridges.

In the middle panels of Fig.~\ref{fig:orbits1filter}, we plot the momentum mapping for orbit 1 deflected by the core with soft collisions. The initial momenta of these orbits [Fig. \ref{fig:orbits1filter}(c)] occupy a much more restricted area around the perpendicular momentum axis and lead to a slightly broader final momentum map, with a v-shaped structure near $(p_z,p_x)=(0,1\;\rm a.u.)$ 
  [see Fig.~\ref{fig:orbits1filter}(d)]. This structure is also present in the single-orbit PMD. 
 Finally, in the bottom panels, we show the mapping for the hard-colliding orbits. Their initial momenta, displayed in Fig.~\ref{fig:orbits1filter}(e), are located in well-defined regions close to the polarization axis, namely a central island similar to that associated with the softly scattered orbits surrounded by two islands much more densely populated around larger parallel momenta. The central island leads to the ridge of energy $10U_p$. In this case, the initial momentum of the particle is so small that the only escape route is to obtain enough kinetic energy via a collision with the core.  The peripheral islands lead to a ridge of much lower energy, associated with a longer return. These ridges are present in the corresponding single-orbit distribution {[Figs.~\ref{fig:single_orbits}(b) and (c)]}. 
 
 One should also note that the initial transverse momenta of the orbits inside the flame in Fig.~\ref{fig:orbits1filter}(a), is around $p_{0x}=0.25 \rm \; a.u$, while the initial momentum of the soft and hard colliding orbits lies below this region. This suggests some kind of cutoff values for the initial transverse momentum, meaning that if the orbits start with lower momenta, they will be more influenced by the core. Consequently, they will have smaller $r_c$, will be detected by our spatial filter and will be classified as colliding orbits. Furthermore, these observations indicate that using the tunnel exit as the spatial filter cannot account for these rare events.

\begin{figure}[!htb]
    \centering
    \includegraphics[width=1\linewidth]{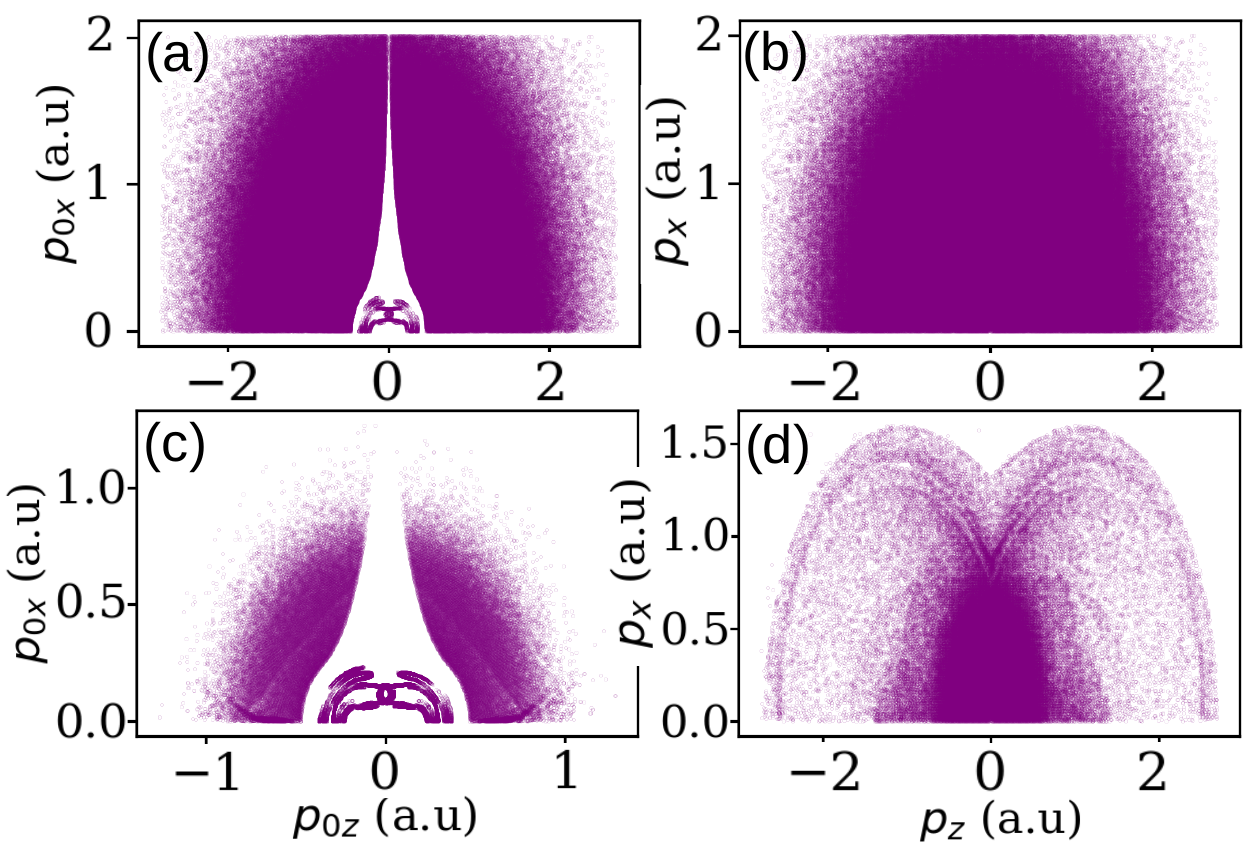}
    \caption{Initial and final momentum of orbit 1, left and right panels, respectively, from a hybrid CQSFA computation using $N=1\times 10^7$ initial orbits. The initial distribution is sampled from the Gaussian distributions $\mathcal{G}_b$ (a) and $\mathcal{G}_n$ (c) which widths are given in Table \ref{tab:width}.}
    \label{fig:orbits1HCQSFA}
\end{figure}

Figure ~\ref{fig:orbits1HCQSFA} shows the initial to final momentum mapping for orbit 1 from the hybrid CQSFA approach, without applying the spatial filtering. The initial conditions were sampled from the broad $\mathcal{G}_b$ [Fig. \ref{fig:orbits1HCQSFA})(a)] and narrow [Fig.~\ref{fig:orbits1HCQSFA})(c)] $\mathcal{G}_n$ Gaussian given in Table ~\ref{tab:width}. For the narrower Gaussian, the initial momentum distribution resembles that obtained with the rate-based approach. This similarity persists in the final momentum distributions, which exhibits a well-defined high energy ridge such as in Fig.~\ref{fig:orbits1filter}(f) and a structure around the perpendicular momentum axis $p_x$ resembling that in Fig.~\ref{fig:orbits1filter}(e). The secondary, low-energy ridge present in Fig.~\ref{fig:orbits1filter}(f) is largely absent, as the narrow Gaussian distribution does not cover sufficient initial momenta in the peripheral islands.  
We can also observe how the density of points in the different momentum regions changes; when using the broad Gaussian, they cover uniformly all the momentum space shown in the plot, while when the width is reduced, they are denser around vanishing momenta. These findings are in agreement with those in Fig.~\ref{fig:arbitrary_Gaussians}, which show that, under a constant number of orbits, narrower initial distributions allow a better probing of the rescatttered trajectories, thus making the associated holographic patterns more resolved. 

\begin{figure}[!htb]
    \centering
    \includegraphics[width=1\linewidth]{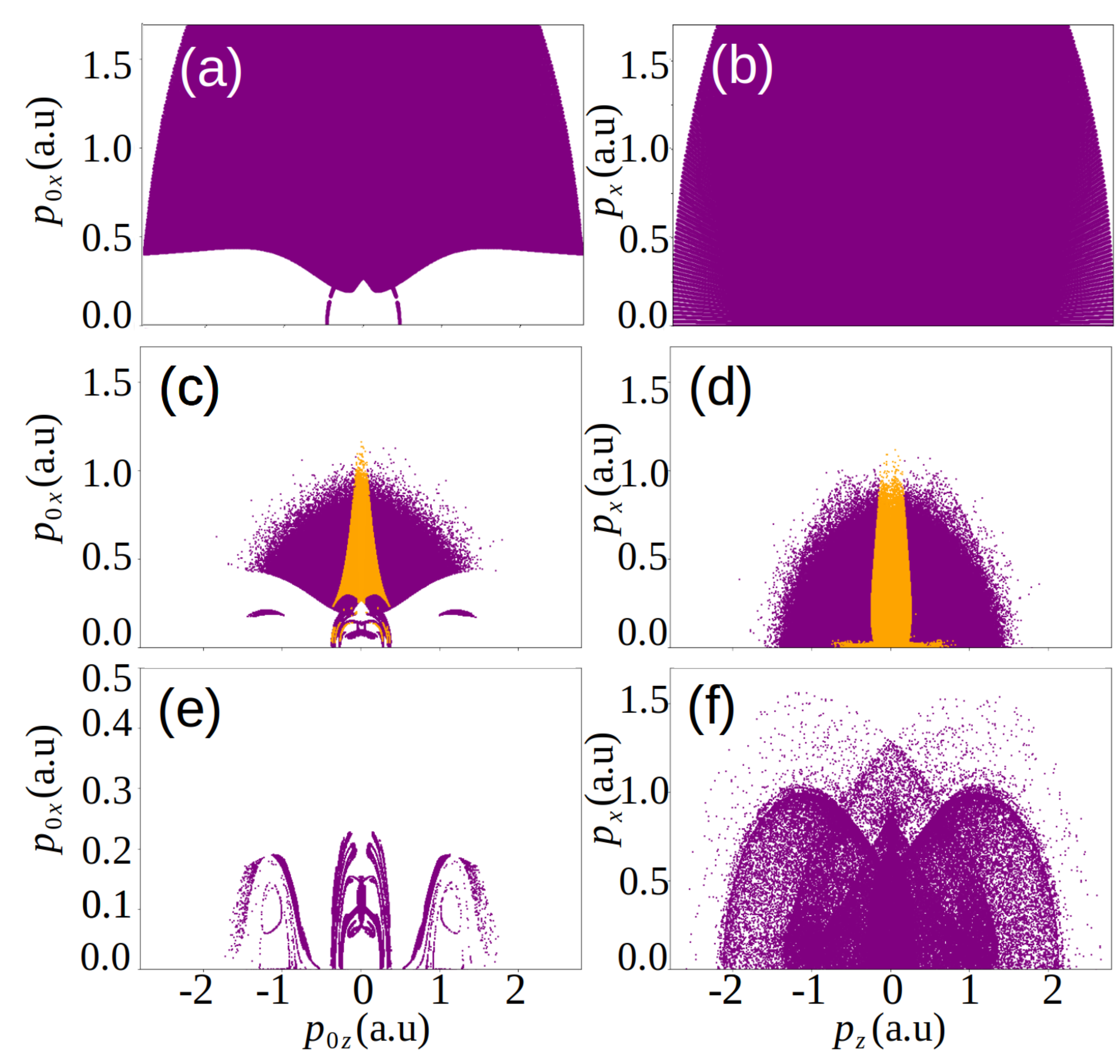}
    \caption{Initial and final momentum of orbit 2, left and right columns, respectively. Panels (a) and (b) show the mapping from a boundary CQSFA calculation. The middle and bottom panels show the mapping from the rate-based method using $N=10^7$ initial orbits. The second row shows non-colliding (orange) and soft colliding orbits (purple). The third row displays the hard-colliding ones.}
    \label{fig:orbits2}
\end{figure}

\subsection{Orbit 2}

In Fig.~\ref{fig:orbits2}, we display the initial (left column) and final (right column) momenta of orbit 2. The top row shows the results obtained within the standard boundary CQSFA framework. The middle and bottom rows show the results from the rate-based method using the spatial filter. The CQSFA momenta, depicted in Fig.~\ref{fig:orbits2}(a), exhibit the behavior expected from field dressed Kepler hyperbolae starting from the ``wrong side" with regard to the detector. At the time of ionization, these orbits need a non-vanishing transverse momentum component to be able to escape with at most a soft collision with the core. For that reason, the bulk of electron trajectories that are classified as orbit 2 have prominent contributions in the large transverse momentum region. This becomes relevant in rate-based or hybrid methods using initially biased distributions, as shown in Fig.~\ref{fig:arbitrary_Gaussians}: if the initial Gaussian distribution is too narrow, the fan may be compromised. 

In panels (c) and (d) we show the superposition of direct (orange) and soft-colliding (purple) orbits according to our classification. The non-colliding orbits 2 are released with non-vanishing transverse momenta and are mainly located along the transverse momentum axis. Their momentum at the detector forms a well-defined structure elongated along the transverse momentum axis, and also a small region along the polarization axis of almost vanishing transverse final momenta. The superposition shown in Fig.~\ref{fig:orbits2}(c) resembles the initial distribution of orbit 2 obtained within the boundary CQSFA approach, except for some structures around vanishing momentum, which is not present in the pure boundary CQSFA calculations. This resemblance is expected, as the vast majority of such orbits are laser-dressed hyperbolae. Furthermore, the orbits type 2 classified as ``direct" are deflected, but their distance of closest approach is never smaller than that defined by the tunnel exit. Thus, the deflection is not picked up by the spatial filter. 

On the other hand, if we consider the hard collision condition, the initial momentum distribution [Fig.~\ref{fig:orbits2}(e)] will be similar to its orbit 1 counterpart. The final momentum distribution [Fig.~\ref{fig:orbits2}(f)] will result in a well-resolved secondary ridge and a caustic. These structures are defined by the central island. The peripheral islands give rise to the structure near the perpendicular momentum axis $p_x$ and remnants of ridges at higher energies. A noteworthy difference between the initial momentum of hard colliding orbits 1 and 2 is observed in the population of the two islands of larger momentum along the polarization axis.  Those from orbit 1 are densely populated and lead to a well-defined rescattering ridge, while those from orbit 2 are sparsely populated. 
\begin{figure}[!htb]
    \centering
    \includegraphics[width=1\linewidth]{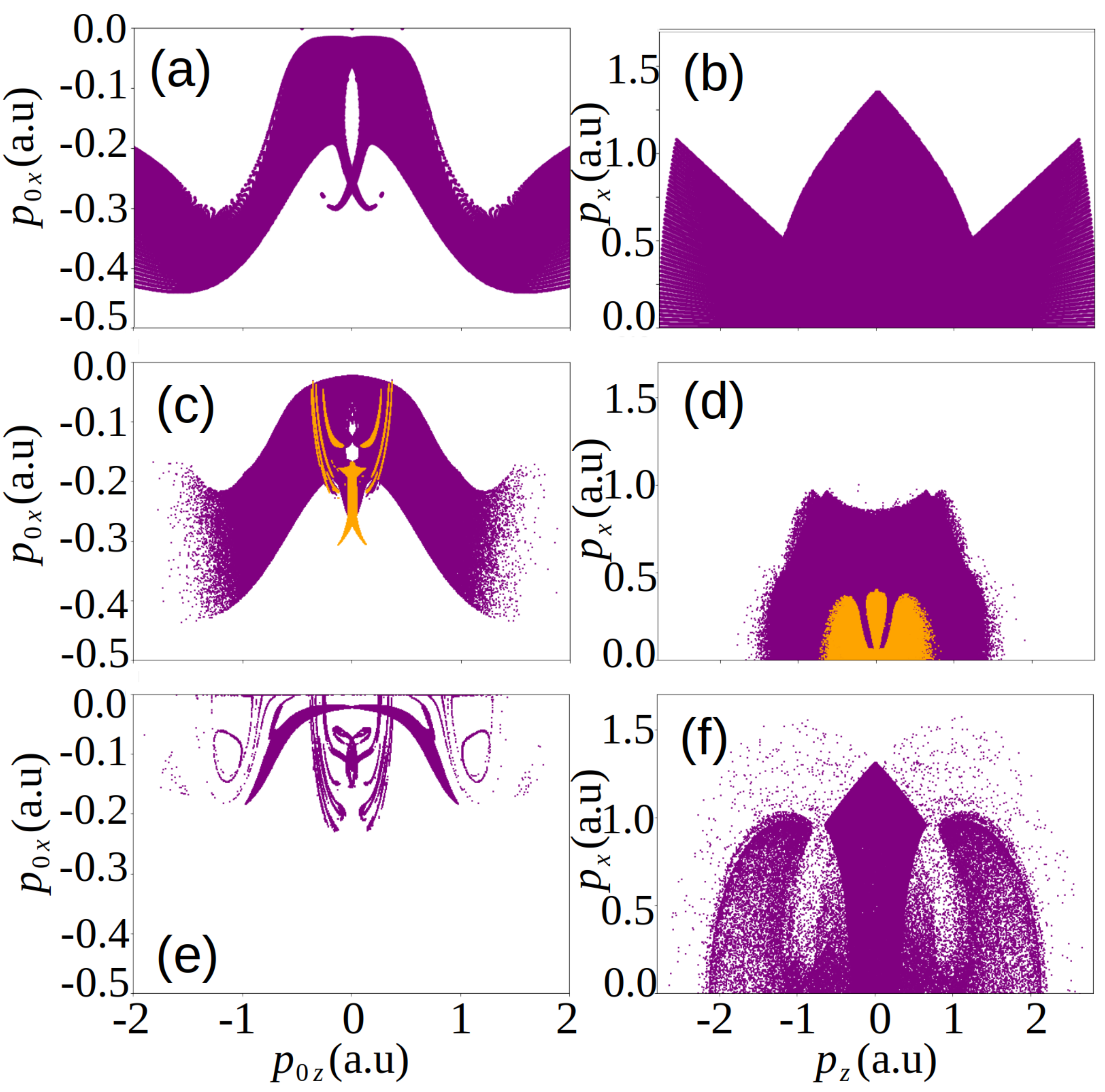}
    \caption{Initial and final momentum of orbit 3, left and right panels, respectively. Panels (a) and (b) show the mapping from a boundary CQSFA calculation. The middle and bottom panels show the mapping from the rate-based method using $N=10^7$ initial orbits. The second row shows non-colliding (orange) and soft colliding orbits (purple). The third row displays the hard-colliding ones.}
    \label{fig:orbits3}
\end{figure}

\subsection{Orbit 3}
Fig.~\ref{fig:orbits3} represents the mapping for orbit 3. As in the previous figures, in the top panel, we show mapping from a boundary CQSFA calculation. The middle panels show direct (orange) and soft-colliding (purple) orbits, and the bottom panel shows the hard-colliding ones, after applying the spatial filter to the results from the rate-based method. An overall feature is that the transverse momentum component changes sign during the electron propagation so that negative values of $p_{0x}$ will map into positive values of $p_x$ and vice versa.  Therefore, even if the spatial filter classifies some orbits 3 as direct, they must still be deflected so that the binding potential will change the transverse momentum in such a way that the conditions in Table \ref{tab:OrbitClassification} hold. 

Interestingly, the initial maps for the boundary CQSFA and the direct and soft-colliding orbits, shown in Figs.~\ref{fig:orbits3}(a) and (c), occupy a much more restricted transverse momentum region than those for the standard orbits 1 and 2, plotted in Figs.~\ref{fig:orbits1}(a) and \ref{fig:orbits2}(a), respectively. Nonetheless, the initial parallel momentum component $p_{0z}$ extends up to relatively large values. This may result in some holographic structures involving orbit 3 being truncated if the initial distribution taken is too narrow (see the example of the spider discussed in Sec.~\ref{sec:sampling} for the Gaussian $\mathcal{G}_n$). The final momentum mapping obtained for the CQSFA [Fig.~\ref{fig:orbits3}(b)] is delimited by a caustic whose apex is located around $(p_z,p_x)=(0,1.3\;\rm a.u.)$ up to perpendicular momenta near $p_x=\pm 0.5 \; \rm a.u. $, but occupies a larger region closer to the field-polarization axis. This region has been shown in our previous publication \cite{Maxwell2018}, and the nature of the orbits changes beyond the caustic.

Next, we will discuss what happens for the rate-based method using the aforementioned spatial filters [Figs.~\ref{fig:orbits3}(c) to (f)]. The initial momentum maps for the direct and soft-recolliding orbits, displayed in Fig.~\ref{fig:orbits3}(c), resembles the CQSFA outcome, while that plotted in Fig.~\ref{fig:orbits3}(e) exhibits a central island and two peripheral islands, all of which are close the field polarization axis. A comparison of the CQSFA and the rate-based method shows that the direct and soft recolliding orbits do not contribute to the caustic [see Figs.~\ref{fig:orbits3}(c) and (d)]. Instead, in both Figs.~\ref{fig:orbits3}(a) and (e) there exists an arch-shaped structure near the origin which unites the two peripheral islands. This structure leads to a caustic in both Figs.~\ref{fig:orbits3}(b) and (f), but not in Fig.~\ref{fig:orbits3}(d). On the other hand, because the direct and soft colliding orbits 3 are present in the boundary-type CQSFA, they fill the structure inside the caustic in Figs.~\ref{fig:orbits3}(b), while there are gaps below the caustic in Fig.~\ref{fig:orbits3}(f). This region being filled can also be seen by inspecting the final momentum map in Fig.~\ref{fig:orbits3}(d), resulting from the initial momenta given in Fig.~\ref{fig:orbits3}(c). 
The central island in Fig.~\ref{fig:orbits3}(e) leads to a rescattering ridge.

Both the direct and soft-colliding orbits exhibit a gap at non-zero transverse momentum. The final momentum distribution of the hard-colliding orbits [Fig.~\ref{fig:orbits3}(f)] resembles the orbit 2 counterpart [Fig.~\ref{fig:orbits2}(f)], except for the empty areas extending along $p_{z}=\pm 1 \rm \;a.u$. This could be understood by comparing the initial momenta of both orbits. We can see how the higher-energy ends of the two islands along the polarization axis are less populated for the hard-rescattered orbit 3 [Fig.~\ref{fig:orbits3}(e)] than for orbit 2 [Fig.~\ref{fig:orbits2}(e)]. Nonetheless, the initial and final momentum maps for the hard colliding orbits 3 resemble their counterparts for orbit 2 [see Fig.~\ref{fig:orbits2}(e) and (f) for comparison], with the difference that the initial momenta lie in the opposite half-plane.

\begin{figure}[!htb]
    \centering
    \includegraphics[width=1\linewidth]{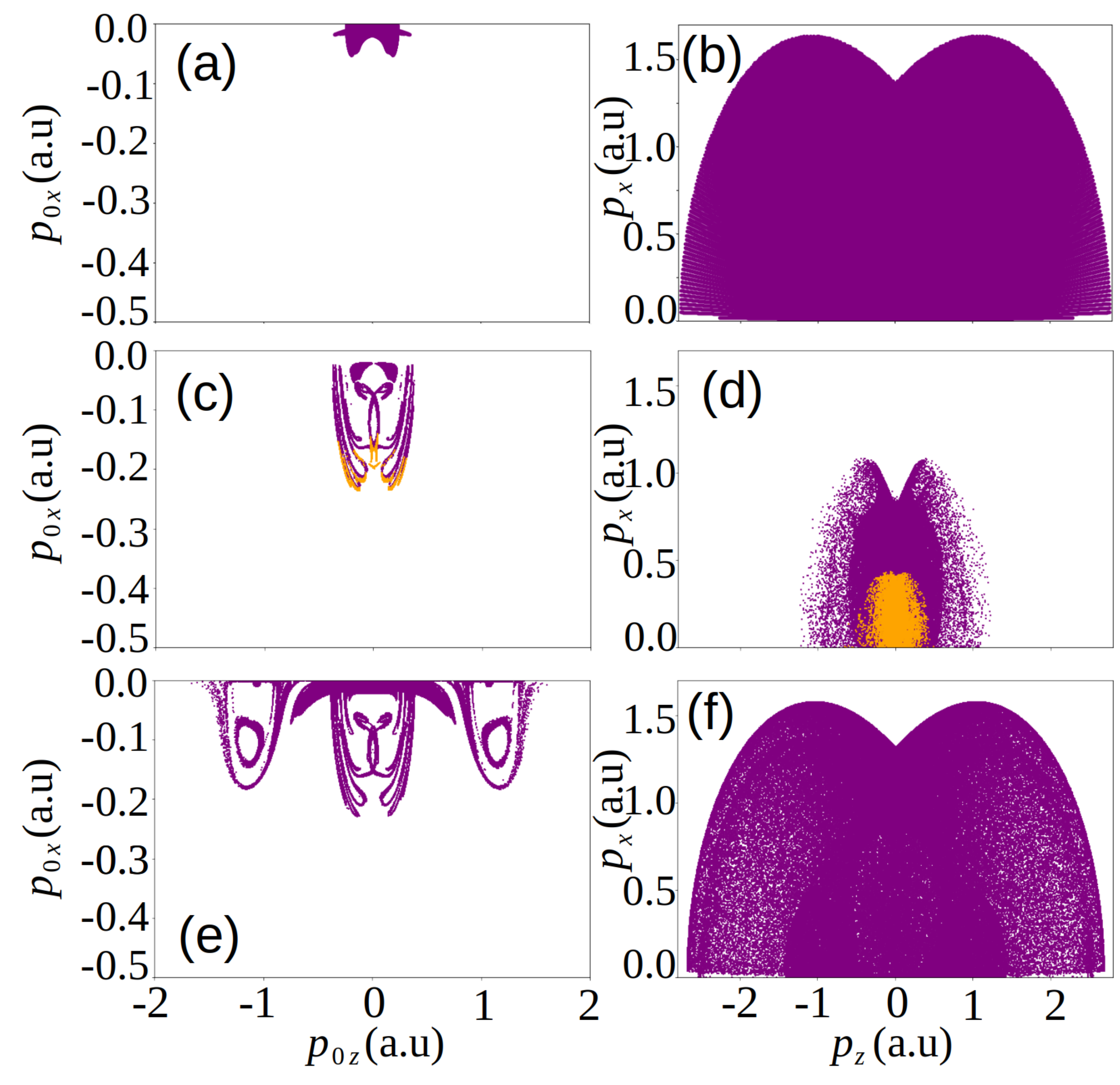}
    \caption{Initial and final momentum of orbit 4, left and right panels, respectively. Panels (a) and (b) show the mapping from a boundary CQSFA calculation. The middle and bottom panels show the mapping from the rate-based method using $N=10^7$ initial orbits. The second row shows non-colliding (orange) and soft colliding orbits (purple). The third row displays the hard-colliding ones.}
    \label{fig:orbits4}
\end{figure}

\subsection{Orbit 4}
Finally, in Fig.~\ref{fig:orbits4}, we plot the initial to final momentum mapping of orbit 4. The top panels stem from a boundary CQSFA calculation, the middle panel shows the direct (orange) and non-colliding (purple) orbits, and the bottom panel shows the hard-colliding orbit 4, after applying the filter to the forward method results. Overall, the initial momentum of these orbits occupies much more restricted momentum regions and low transverse momenta, in the opposite half plane of the final momenta. This shows that orbit 4 starts close to the field polarization axis. 

The standard CQSFA outcome shows an initial momentum region localized in the vicinity of the origin ($p_{0z},p_{0x})=(0,0)$ [Fig.~\ref{fig:orbits4}(a)], which leads to a large final momentum region whose boundary is the high-energy rescattering ridge [Fig.~\ref{fig:orbits4}(b)]. In contrast, the mapping from the rate-based method includes other types of orbit 4. The initial momenta of the non-colliding orbits within our classification [Fig.~\ref{fig:orbits4}(c)] occupy a small region located around $p_{0x}=-0.2 \rm \;a.u$ and $(-0.5 \rm\; a.u.< p_{0z}<0.5\rm \; a.u.)$, reaching the detector with small momentum values located along the transverse momentum axis. The initial momenta of soft-colliding orbits are also localized within the same region along the parallel axis $(-0.5\rm\; a.u.<p_{0z}<0.5\rm \; a.u.)$, but extend longer in the perpendicular direction. The initial momentum distribution of the hard-colliding orbits resembles the distribution encountered for all the other hard-colliding orbits, meaning a central island, and two islands of higher parallel initial momentum. The final momentum distribution exhibits a high energy rescattering ridge, typically expected for these orbits, as shown in the final momentum distribution from the boundary CQSFA calculation. Comparing the initial momenta displayed in Fig.~\ref{fig:orbits4}(a) and Fig.~\ref{fig:orbits4}(e) we observe that only a small region around vanishing initial momentum is obtained with the boundary CQSFA calculations, and this is the one already leading to the high energy rescattering ridge. The missing orbits in the boundary CQSFA are the reason behind the absence of many holographic patterns that are observed for the rate-based and hybrid approaches associated solely with orbit 4, such as the annular interference structures following the primary rescattering ridge. 

Interestingly, there is a gap around the vanishing initial transverse momentum for all orbits except orbit 4. We see this, both with the rate-based method and the boundary CQSFA approaches. Also, we observe clear high-energy rescattering ridges, both for hard-scattered orbits 1 and 4. Looking at the dynamics of two of the orbits leading to this structure, which end up classified as orbits 1 and 4, reveals that both start with small transverse momentum but with opposite signs, and orbit 1 undergoes hard collisions with the core later on, as they are driven by the field before experiencing hard collisions. 

Furthermore, the previous analysis allows us to understand why the effect of the Coulomb correction added to the ionization rate was more noticeable on the single PMDs of orbits 1 and 2. From the momentum mapping, we can see how the initial momentum of non-colliding orbit 1, the one that most resembles the expected behavior of direct orbits, and the soft and non-colliding orbit 2 extend to higher values along the transverse momentum axis, being more sensitive to the changes of the width of the distribution in this direction. This also explains our observations regarding  the extension of the fan, which originates from the interference of direct orbits and field-dressed hyperbolae.

\section{Conclusions}
\label{sec:conclusions}
In the present work, we develop forward and hybrid methods that are orbit-based, Coulomb distorted, and carry tunneling and quantum interference, and apply them to ultrafast photoelectron holography. These methods are more versatile than the Coulomb quantum-orbit strong-field approximation (CQSFA) in its original form, as they do not rely on any pre-knowledge or assumption about the dynamics of the contributing orbits. Although they use elements of the CQSFA, they start by launching an ensemble of Coulomb-distorted trajectories that are either used as guesses for a boundary problem, or propagated up to their asymptotic value. The former strategy is employed in a hybrid CQSFA (H-CQSFA) and the latter in a forward rate-based method. In contrast, the CQSFA in its original form is implemented as a boundary problem in which specific assumptions upon the orbits are made, the standard, Coulomb-free SFA is used as a first approximation and the binding potential is introduced incrementally. The methods agree well with the TDSE computations and among each other, and additional features are encountered, in comparison with the standard CQSFA, such as annular interference patterns and additional rescattering ridges. The main holographic structures, such as the fan, the spider and the spiral, are present in all cases.

A key finding is that both proposed methods are strongly dependent on the initial conditions at the tunnel exit, which are either incorporated as ionization rates or sub-barrier contributions to the semiclassical action. The ionization rate employed in the present article was constructed using the sub-barrier contributions of the CQSFA in the low-frequency approximation. Therefore, besides being non-adiabatic, similar to the one encountered in \cite{Li2016PRA} it exhibits Coulomb corrections. Our results prove the relevance of including the latter in the rate by achieving a better agreement with the hybrid CQSFA calculations. In a previous publication, sub-barrier corrections have been used to construct analytic approximations for the CQSFA \cite{Maxwell2017a}. However, this was done in a boundary-type problem instead of being used in a rate. In our rate-based method, once the trajectories are sampled from a given ionization probability, they will carry this weight along the propagation, and it will contribute to the final PMDs.
The hybrid CQSFA  also launches a set of initial conditions, but, as the boundary problem is solved, the effect of sampling from different distributions will not affect the weight of the trajectories at the detector. However, it can impact the stability of the solver, or for a fixed number of trajectories, it can change how we resolve the different holographic patterns.

Examples of this influence have been provided in Sec.~\ref{sec:PMDs}. Therein, it was shown that the sub-barrier Coulomb corrections broaden (narrow) the initial electron-momentum distribution in the direction perpendicular (parallel) to the laser-field polarization. This will influence holographic patterns such as the fan, the spider and the carpet  
(see Sec.~\ref{sec:singleorbit}). Furthermore, by choosing an arbitrarily narrow or broad initial sampling, one may emphasize or leave out groups of orbits and their contributions to the specific holographic structures. For instance, a narrow Gaussian will produce better rescattering ridges and interference features in that region, but will result in a coarser fan and a truncated spider. These outcomes are related to the orbits whose interference leads to the structures: those released in the continuum near the polarization axis will be favored by narrower initial sampling, while those freed with large initial transverse momenta will be probed better if the initial conditions are sampled broadly. 
 
In addition, care must be taken with pre-assuming the orbits' dynamics. The conditions upon the tunnel exit and momentum components used in their classification and given in Table \ref{tab:OrbitClassification} are insufficient to guarantee the behavior first stated in \cite{Yan2010} and used in the original, boundary-type CQSFA \cite{Lai2015a,Lai2017,Maxwell2017}.  The underlying assumptions that orbit 1 goes directly to the detector, orbits 2 and 3 are laser-dressed hyperbolae and orbit 4 goes around the core before reaching the detector leaves out whole classes of orbits, which behave differently but still fulfill the conditions in Table \ref{tab:OrbitClassification}. Although these orbits do not influence standard holographic structures, such as the fan, the spider or the spiral, they lead to additional rescattering ridges, low-energy structures and caustics, which appear both in momentum mappings or single-orbit distributions. Furthermore, their interference may lead to additional holographic structures. Examples are the interference of different types of orbit 4, which can be associated with pairs of short and long rescattered orbits that exist in the Coulomb free strong-field approximation \cite{Paulus1994,Paulus1994JPhysB,Faria2002}, fork-like structures and secondary ridges \cite{Moeller2014,Becker2015}, and also additional types of orbits in the spiral. These latter orbits have been identified recently as multi-pass trajectories \cite{Qin2021}. 

A further study also calls into question the nature of the orbits. In the boundary problems solved by us so far, orbits 1 and 2 are direct or at most lightly deflected by potential, orbit 3 is a hybrid, and orbit 4 is rescattered. However, single-orbit distributions have revealed rescattering ridges for orbits 1, 2 and 3 as well. For that reason, we have employed the spatial filter from our previous publication \cite{Maxwell2018} to sort the orbits launched by the forward and hybrid approaches into ``direct'', ``deflected" and ``hard scattered". Comparing the orbits' distance of closest approach $r_c$ with the tunnel exit $|z_0|$ and the Bohr radius $r_0$, we have called direct orbits all those trajectories for which $r_c\geq |z_0| $, soft scattered those for which $r_0 < r_c< |z_0| $ and hard scattered if $r_c$ is smaller or equal to the Bohr radius. These assumptions have resulted in all types of CQSFA orbits, from 1 to 4, having direct, soft scattered, and hard scattered subsets. Directs orbits 1, and soft scattered orbits 2 and 3 correspond broadly to those obtained with the standard CQSFA. Orbit 4 in the standard CQSFA corresponds to a subset of the hard scattered orbits 4 for the forward and hybrid methods. 

We have explored these types of orbits in detail in initial to final momentum maps, which revealed several noteworthy features. First, all hard scattered orbits lead to three islands along the polarization axis in the $p_{0x}p_{0z}$ plane for the initial momenta: a central island near the origin and two peripheral islands centered at non-vanishing parallel momenta. For the final momenta, the central island leads to rescattering ridges, whose energy depends on the orbit in question, and the peripheral islands lead to caustics and low-energy structures. The other sets of orbits behave in a less universal way regarding the initial conditions, but there are similarities. In general, direct orbits fill the whole final momentum grid, while soft-scattered orbits fill momentum regions in the vicinity of $p_z=0$ extending to relatively large perpendicular final momentum. Particularly similar is the momentum mapping of the soft-scattered orbits 1 and 4, which start from a central island and fill a momentum region near the perpendicular momentum axis.  

One should note, however, that spatial filtering using the distance to the core brings some degree of arbitrariness. For instance, there are deflected orbits that are classified as ``direct" because their perihelion is larger than or equal to the distance defined by the tunnel exit. This happens to orbits 2, 3 and in particular 4. For orbit 1, there is also a subset of orbits that start inside the flame-shaped gap and are deflected but not detected by the spatial filter.
Still, it is remarkable how neat the outcome of the filtering is, in general. Other types of criteria have been used in the literature to separate soft from hard collisions, such as, for instance, the scattering angle \cite{Kelvich2016}. Finally, the method developed in this work is flexible enough to be applied to photoelectron holography in tailored fields. Additionally, it enables an in-depth study of caustics, ridges and low-energy structures in a fully Coulomb-distorted framework incorporating tunneling and quantum interference, or may be extended to scenarios with more than one active electron.

\begin{acknowledgments}

We thank C. Hofmann, N. Shvetsov-Shilovsky ans S. Brennecke for useful discussions. This work was funded by grants No.\ EP/J019143/1 and EP/T517793/1, from the UK Engineering and Physical Sciences Research Council (EPSRC).  A.S.M. acknowledges funding support from
the European Union’s Horizon 2020 research and innovation program under the Marie Sk{\l}odowska-Curie Grant
Agreement SSFI No. 887153.
\end{acknowledgments}

\bibliography{Review2.bib}

\end{document}